\documentclass{article}

\usepackage{arxiv}

\usepackage[utf8]{inputenc} 
\usepackage[T1]{fontenc}    
\usepackage{hyperref}       
\usepackage{url}            
\usepackage{booktabs}       
\usepackage{amsfonts}       
\usepackage{nicefrac}       
\usepackage{microtype}      
\usepackage{xcolor}         

\usepackage{comment}
\usepackage{todonotes}
\usepackage{graphicx}
\usepackage{subfigure}
\usepackage{bbm}
\usepackage{mathtools}
\usepackage{verbatim}
\usepackage[linesnumbered, ruled, vlined]{algorithm2e}
\usepackage{enumitem}
\usepackage{breqn} 

\usepackage{hyperref}
\usepackage{wrapfig}
\usepackage{soul} 

\usepackage{longtable} 

\usepackage{multirow}

\usepackage{upgreek}

\usepackage{gensymb} 

\usepackage[capitalize,nameinlink]{cleveref}
\usepackage[title]{appendix}
\crefname{appsec}{appendix}{appendices}
\Crefname{appsec}{Appendix}{Appendices}

\usepackage{tikz}
\usetikzlibrary{intersections,shapes.arrows}

\definecolor{richardcolor}{rgb}{0.2, 0.4, 0.0}

\usepackage{amsthm,amsmath,amssymb,amsfonts}

\allowdisplaybreaks

\usepackage{amsmath,amsfonts,bm}
\usepackage{dsfont}

\usepackage{amsmath,amsfonts,bm}

\def\eqref#1{equation~\ref{#1}}

\def\1{\bm{1}}

\DeclareMathAlphabet{\mathsfit}{\encodingdefault}{\sfdefault}{m}{sl}
\SetMathAlphabet{\mathsfit}{bold}{\encodingdefault}{\sfdefault}{bx}{n}

\usepackage{hyperref}

\title{Few-Shot Learning Patterns in Financial Time-Series for Trend-Following Strategies}

\author{%
  Kieran Wood$^*$\\
  Oxford-Man Institute\\
  University of Oxford\\
  \texttt{kieran.wood@eng.ox.ac.uk} \\
  \And
  Samuel Kessler$^*$\\
  Oxford-Man Institute\\
  University of Oxford\\
  \texttt{samuel.kessler@eng.ox.ac.uk} \\
  \AND
  Stephen J. Roberts \\
  Oxford-Man Institute\\
  University of Oxford\\
  \texttt{stephen.roberts@eng.ox.ac.uk} \\
  \And
  Stefan Zohren \\
  Oxford-Man Institute\\
  University of Oxford\\
  \texttt{stefan.zohren@eng.ox.ac.uk} \\
}

\renewcommand{\shorttitle}{Few-Shot Learning Patterns in Financial Time-Series for Trend-Following Strategies}

\hypersetup{
pdftitle={Few-Shot Learning Patterns in Financial Time-Series for Trend-Following Strategies},
pdfsubject={q-fin.TR, cs.LG, stat.ML},
pdfauthor={Kieran~Wood, Samuel~Kessler, Stephen J.~Roberts, Stefan~Zohren},
pdfkeywords={Trend-Following, Time-Series Momentum, Few-Shot Learning, Deep Learning, Machine Learning, Transfer Learning, Change-point Detection, Quantitative Finance, Systematic Strategies, Portfolio Construction},
}

\begin{document}

\maketitle
\def\thefootnote{*}\footnotetext{Equal contribution.}\def\thefootnote{\arabic{footnote}}
\begin{abstract}
Forecasting models for systematic trading strategies do not adapt quickly when financial market conditions rapidly change, as was seen in the advent of the COVID-19 pandemic in $2020$, causing many forecasting models to take loss-making positions. To deal with such situations, we propose a novel time-series trend-following forecaster that can quickly adapt to new market conditions, referred to as \emph{regimes}. We leverage recent developments from the deep learning community and use \emph{few-shot} learning. We propose the \textbf{Cross} Attentive Time-Series \textbf{Trend} Network -- \textbf{X-Trend} -- which takes positions attending over a context set of financial time-series regimes. X-Trend transfers trends from similar patterns in the context set to make forecasts, then subsequently takes positions for a new distinct target regime. By quickly adapting to new financial regimes, X-Trend increases Sharpe ratio by $18.9\%$ over a neural forecaster and 10-fold over a conventional \emph{Time-series Momentum} strategy during the turbulent market period from $2018$ to $2023$. Our strategy recovers twice as quickly from the COVID-19 drawdown compared to the neural-forecaster. X-Trend can also take \emph{zero-shot} positions on novel unseen financial assets obtaining a 5-fold Sharpe ratio increase versus a neural time-series trend forecaster over the same period. Furthermore, the cross-attention mechanism allows us to interpret the relationship between forecasts and patterns in the context set.
\end{abstract}

\keywords{Trend-Following \and Time-Series Momentum \and Few-Shot Learning \and Deep Learning \and Machine Learning \and Transfer Learning \and Change-point Detection \and Quantitative Finance \and Portfolio Construction}

\section{Introduction}
When financial market conditions change the forecasting models that are used to take positions in the markets perform very poorly~\cite{MomentumCrashes, MomentumTurningPoints}. The recent success of deep learning for learning representations from data has translated into better financial forecasting models~\cite{lim2019enhancing}. However deep learning models also rely on large stationary datasets for representation learning. Financial markets can be highly non-stationary due to changing market conditions. When a financial market enters a new \emph{regime}, augmenting the inputs with indicators of the time and severity of regime change improves returns~\cite{wood2022slow}. This finding is important since it shows that there is a benefit to training deep learning models that have additional supervision about when regimes change. Furthermore, it raises an important question: can we selectively use historical patterns to transfer knowledge of the past to make forecasts for new regimes and new markets?
 
\begin{figure}[h]
    \centering
    \includegraphics[width=0.9\textwidth]
    {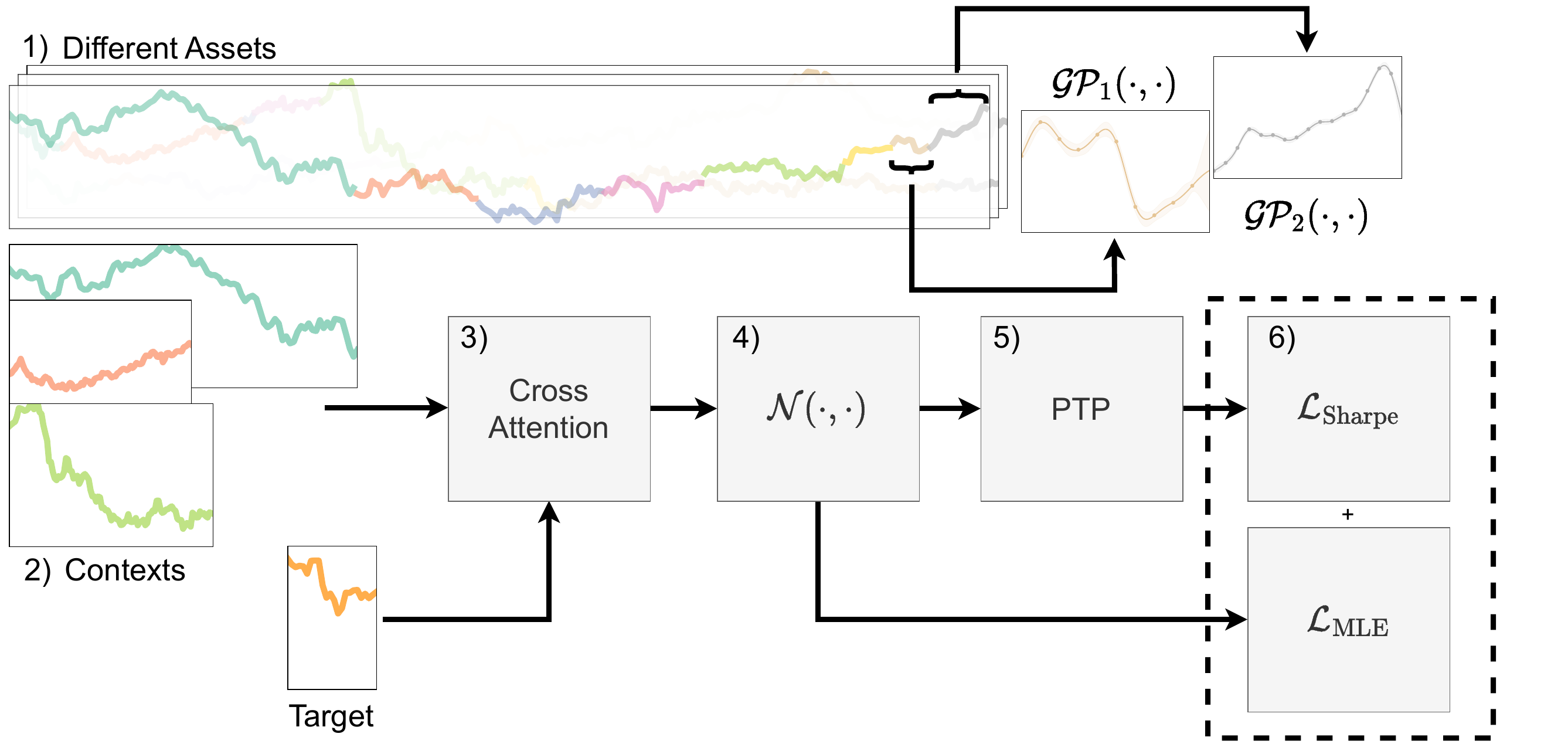}
    \caption{An overview of the X-Trend few-shot learning trend-following model. 1) Each asset is segmented into regimes using a change-point detection algorithm. 2) The context set is constructed by randomly sampling regimes from different assets. The objective is to produce long/short positions given the target sequence while respecting causality with the context set. 3) Our X-Trend model uses a cross-attention layer for the target to leverage patterns in the context set. 4) The model produces a distribution over next-day returns. 5) The model outputs positions using \textbf{P}redictive probability density function \textbf{T}o traded \textbf{P}osition module (PTP). 6) We train our model by jointly optimizing a Sharpe ratio loss and negative log-likelihood.}
    \label{fig:flow-diagram}
\end{figure}

In recent years it has been observed, that the risk-adjusted returns of conventional \emph{Time-series Momentum} (TSMOM) models~\cite{moskowitz2012time}, which exploit trends in financial time-series, have deteriorated by $87.4\%$ from $2018$ to $2023$ compared to the period from $1995$ to $2000$. This can potentially be attributed to a concept known as `factor crowding'~\cite{FactorCrowding}, where arbitrageurs trade the same assets based on similar factors. This causes market inefficiencies to quickly disappear and can increase the risk of liquidity-driven tail events~\cite{brown2022crowded}. To mitigate this, one alternative has been to explore new assets where we typically do not have sufficient data for deep learning approaches. 

In deep learning, quick adaptation to new data or \emph{few-shot learning} has seen recent advances in computer vision \cite{vinyals2016matching, ravi2016optimization, snell2017prototypical}. Few-shot learning involves training neural networks (NNs) such that they are able to adapt and learn from minimal data. Few-shot learners are tested on completely unseen classes of images using a few to no examples or are required to solve unseen reinforcement learning environments~\cite{duan2016rl, wang2016learning, finn2017model}. Few-shot learners have the desirable quality of being able to adapt and learn from very few data points. This is advantageous for a systematic trading strategy, allowing it to adapt quickly to new financial regimes or new markets. Broadly, we can categorize the market regimes, which systematic strategies aim to exploit, as \emph{trending} or \emph{mean-reverting}. Mean-reversion is a market phenomenon whereby, under certain circumstances, the price has a tendency to return its long-term mean~\cite{poterba1988mean}. A detailed study of the \emph{trending} and \emph{mean-reverting} financial market anomalies can be found in~\cite{vayanos2013institutional}.

In this work, we leverage advances in few-shot learning and time-series momentum strategies to develop a model that can make predictions in new market regimes and unseen markets. In practice, our experiments backtest on continuous futures contracts of various asset classes: equities, foreign exchange, commodities and fixed income. Our model obtains significantly higher risk-adjusted returns in terms of Sharpe ratio~\cite{SharpeRatio}, a measure of returns per unit volatility. Our model is able to learn transferable patterns, then subsequently learn to take positions in markets or regimes distinct from those used for training. The idea of universal patterns in financial markets, which are transferable, is motivated by the works ~\cite{RamaContUniversality, zhang2019deeplob, wood2021trading}. 

Our model uses a cross-attention mechanism over a context-set \cite{kim2019attentive, doersch2020crosstransformers}. Attention mechanisms~\cite{vaswani2017attention} have been shown to improve returns by attending over an asset's history~\cite{wood2021trading}, this work generalizes this finding by extending the temporal attention mechanism over other assets. This enables our model to transfer knowledge from the context set to enable better predictions for a new regime or new market with little data. Different regimes from financial assets are segmented using change-point detection methods~\cite{garnett2010sequential, saatcci2010gaussian}. 
Additionally, the attention maps provide a degree of interpretability in the resulting predictions~\cite{TFT, wood2021trading}. We also use the latest insights from deep time-series momentum strategies to train our model to produce positive returns over our baselines~\cite{lim2019enhancing}. We call our method the \textbf{Cross} Attentive Time-Series \textbf{Trend} Network or \textbf{X-Trend} for short. The code is available at \href{https://github.com/kieranjwood/x-trend}{https://github.com/kieranjwood/x-trend}. 

We summarize our contributions as follows:
\begin{itemize}
    \item We leverage few-shot learning, and change-point detection to develop an agent which is able to produce returns in futures markets with minimal data. Our X-Trend model is able to successfully respond to ``momentum crashes''~\cite{MomentumCrashes}, and ``momentum turning points''~\cite{MomentumTurningPoints}. We improve Sharpe ratio by $18.9\%$ in comparison to the benchmark neural forecaster over $2018$ to $2023$, an extremely turbulent period in various financial markets. Our X-Trend strategy recovers from the initial COVID-19 drawdown twice as quickly.
    \item X-Trend learns to make predictions in challenging, low-resource, zero-shot settings where the model has never seen a financial asset during training. It improves over the loss-making neural forecaster to achieve an average Sharpe of 0.47 over the period $2018$ to $2023$. This turbulent period is particularly challenging for unseen financial assets.
    \item X-Trend makes interpretable predictions. It is able to learn relationships between similar assets using an interpretable cross-attention mechanism over a context set of different assets. For a given target sequence, a similarity score with patterns in the context set via cross-attention can be visualized. Additionally, by outputting a forecast as an auxiliary output step, we reveal the relationship between optimal trading signal and forecast.
\end{itemize}

\section{Preliminaries}
\label{sec:prelims}
Let us denote a time-series $p^{(i)}_{1:t}$ of daily close prices, where $i$ denotes a particular asset from a basket of assets $i \in \mathcal{I}$ and  $t$ denotes a time index $t \in \{1,\ldots ,T\}$ where $T$ is the final observation for the asset. We work with returns $r^{(i)}_{1:t}$, which linearly de-trends the price series:
\begin{equation}
    r_{t-t',t}^{(i)} = \frac{p_t^{(i)} - p_{t-t'}^{(i)}}{p_{t-t'}^{(i)}},
\end{equation}
where $t'$ is the number of days we calculate returns over and for brevity we use $r_{t}^{(i)}$ to denote $r_{t-1,t}^{(i)}$. 

Our objective is to trade a position $z^{(i)}_{t} \in [-1, 1]$ which we hold for the next day $t+1$ conditioned on a \emph{target} sequence $\mathbf{x}^{(i)}_{-l_t:t}$ of the $l_t$ last days, where $\mathbf{x}^{(i)}_t\in\mathcal{X}$ is a vector of $|\mathcal{X}|$ factors measuring trends at different timescales or the relationship between trends at different timescales. These factors are constructed for time $t$ using returns data and price data, which we elaborate on in~\Cref{sec:classical_approaches} and~\Cref{sec:mom-deep-learning}. We aim to choose positions such that we maximize portfolio returns:
\begin{equation}
\label{eqn:tsmom}
R_{t+1}^{\text{Port.}} = \frac{1}{N} \sum_{i=1}^{N} R_{t+1}^{(i)}, \quad \text{where } R_{t+1}^{(i)} = z_t^{(i)}~\frac{\sigma_{\mathrm{tgt}}}{\sigma_t^{(i)}}~r_{t+1}^{(i)} - C^{(i)}~\sigma_{\mathrm{tgt}} \left| \frac{ z_t^{(i)}}{\sigma_t^{(i)}} - \frac{ z_{t-1}^{(i)}}{\sigma_{t-1}^{(i)}} \right|,
\end{equation}
for each of the assets $i \in \mathcal{I}$ where $N = |\mathcal{I}|$ assets and $C^{(i)}$ is the transaction cost. We use \emph{volatility targeting}~\cite{moskowitz2012time, TSMomAndVolScaling, VolTargeting}, which introduces a leverage factor $\sigma_{\mathrm{tgt}}/\sigma_t^{(i)}$, where we normalize our holdings by the ex-ante volatility, $\sigma_t^{(i)}$, then scale by the annual target volatility $\sigma_{\mathrm{tgt}}$. Aligning our work with the literature~\cite{moskowitz2012time, TSMomAndVolScaling}, $\sigma_t^{(i)}$ is calculated using an exponentially weighted moving standard deviation of returns over the span $r_{-60:t}$, where the contribution has decreased to zero by 60 days in the past. For this paper we set $C^{(i)}=0$ rather than assuming a cost for each asset, focusing on pure predictive power of our model. The volatility targeting approach to portfolio construction ensures that each asset contributes approximately equal risk to the portfolio\footnote{We assume a diagonal covariance, which is a reasonable assumption outside of the tail for the basket of futures contracts we consider in this paper, which is a typical basket for trend-following strategies.}. As such, we perform regression to estimate the probability distribution of volatility scaled next-day returns, $~\frac{\sigma_{\mathrm{tgt}}}{\sigma_t^{(i)}}~r_{t+1}^{(i)}\in \mathbb{R}$.

\subsection{Episodic Learning}
\label{sec:episodic_learning}
We use \emph{episodic learning}~\cite{vinyals2016matching} which trains models in the same way as they are used for testing; models are trained to produce few-shot and zero-shot predictions. Traditional deep learning puts
 data from all assets together and is trained using mini-batch stochastic gradient descent~\cite{robbins1951stochastic, DeepLearningBook}. In episodic learning, we want to make forecasts given a sequence's history for a specific asset, and leverage sequences from other assets for additional non-parametric similarity learning.

Our learner selects a position given \emph{target}: $\mathbf{x}^{(i)}_{t - l_t + 1:t}$, which we shorten to $\mathbf{x}^{(i)}_{-l_t:t}$ for brevity, and where $i$ belongs to our test set $\mathcal{I}_{ts}$ of assets. Additionally, the learner leverages a \emph{context} set of prices $\mathcal{C} = \left\{ \mathbf{x}^{(c)}_{-l_c:t_{c}} \right\}_{c = 1}^{C}$ where $c$ belongs to our training set of assets $\mathcal{I}_{tr}$, $l_{c}$ is the length of each time-series in the context set and $C = |\mathcal{C}|$ is the total number of contexts in the context set. In all circumstances, the context sets are time-series that have occurred before the time-series in the target: $t_{c}  <  t$ for all $c$ in the context set so that our predictions are causal. Throughout the paper we will work with two distinct problem scenarios:
\begin{itemize}
    \item \emph{Few-shot}: where the context set can contain the same asset as the target (but in the past) $\mathcal{I}_{tr} \cap \mathcal{I}_{ts} = \mathcal{I}$.
    \item \emph{Zero-shot}: where the target set asset (at test time) is not contained in the context set $\mathcal{I}_{tr} \cap \mathcal{I}_{ts} = \emptyset$.
\end{itemize}
We sample a target time-series, $\mathbf{x}_{-l_t:t}^{(i)}$ and we sample an associated context set $\mathcal{C}$ per target point in the mini-batch. As opposed to our target problem, we also include the volatility scaled next-day returns as inputs in the context set, denoting the combined inputs as $\bm{\xi}^{(c)}_{-l_c:t}$. We describe how we construct our context sets for the financial assets in~\Cref{sec:context_set_construction}.

\subsection{Classical Momentum Approaches}
\label{sec:classical_approaches}
Time-series Momentum (TSMOM) strategies~\cite{moskowitz2012time}, or \emph{trend-following} strategies, are based on the idea that strong price trends have a tendency to persist. The phenomena of trend-following has been observed historically for more than a century\footnote{It should be noted that returns have started to suffer since the introduction of electronic trading~\cite{wood2022slow} this century.}, outperforming a simple buy-and-hold \emph{Long} strategy~\cite{CenturyOfTrendFollowing}. TSMOM strategies aim to forecast trends and then map them to a trading signal, or position. For instance, we can calculate returns over the past year ($252$ trading days), $r^{(i)}_{t-252,t}$ and map this to a position $z_t^{(i)}$:
\begin{equation}
    z_t^{(i)} = \mathrm{sgn}(r^{(i)}_{t-252,t}),
\end{equation}
where $\mathrm{sgn}(\cdot)$ is the sign function, with $+1$ corresponding to a full long position and $-1$ a full short position~\cite{moskowitz2012time}.  The \emph{Long} strategy takes a full long position at each time-step: $z_t^{(i)} = 1$.

We employ volatility scaling for portfolio construction~\cref{eqn:tsmom}. This approach to portfolio construction is often used in practice by Commodity Trading Advisors (CTAs), where trend-following is a major component of their overall strategy. While this TSMOM approach has proven to be successful over time~\cite{CenturyOfTrendFollowing}, the $1$-year momentum indicator particularly suffers during periods where market conditions change rapidly. We call these \emph{regime changes} throughout this paper (also referred to as momentum crashes~\cite{MomentumCrashes}). An attempt to mitigate regime changes is to combine weighted signals at different timescales, for example, 1-month (21-day), 3-month (63-day), half-year (126-day) and 12-month momentum, $z_t^{(i)}=\sum_{t'\in\{21, 63, 126, 252\}}w_{t'}~\mathrm{sgn}(r^{(i)}_{t-t',t})$, where $w_{t'}\in[0,1]$ represents the weighting of each respective factor. 

MACD (Moving Average Convergence Divergence) factors compare exponentially weighted signals at two different timescales~\cite{AHLMomentum}. These popular momentum indicators aim to balance the trade-off between capturing trends and responding to potential regime changes:
\begin{subequations}
\begin{align}
    \mathrm{MACD}\left(p_{1:t}^{(i)}, S, L\right) &= \frac{m^{(i)}_t}{\mathrm{std}\left(m_{-252:t}^{(i)}\right)} \\
    \textrm{where} \, m_t &= \frac{\mathrm{EWMA}\left(p_{1:t}^{(i)}, S\right) - \mathrm{EWMA}\left(p_{1:t}^{(i)}, L\right)}{\mathrm{std}\left(p_{-60:t}^{(i)}\right)},
\end{align}
\end{subequations}
where $\mathrm{EWMA}(\cdot)$ function is an exponentially weighted moving average. The inputs are a short timescale $S$, with half-life of $\log \left(\frac{1}{2}\right) / \log \left(1 - \frac{1}{S}\right)$ and a long timescale $L$, defined similarly. The MACD signal indicates buy if $>0$ and sell if $<0$. The magnitude provides a measure of conviction or signal strength. It is common to blend multiple MACD indicators at different timescales with a typical choice being $(S,L) \in \left\{(8,24), (16,28), (32,96)\right\}$~\cite{AHLMomentum}. Funds typically convert the MACD signal to a position via response function: $y \mapsto y~\exp(-y^2/4)/0.89$~\cite{AHLMomentum}.

\subsection{Deep-learning Momentum Approaches}
\label{sec:mom-deep-learning}
In financial markets, we often observe different trends and mean-reversions, which we can also think of as a collection of shorter-term trends, occurring concurrently at multiple timescales. Furthermore, the added complexity of regime change means that it is a daunting task to successfully blend various trading signals. This motivated \emph{Deep Momentum Networks} (DMNs), as a solution to such a complex forecasting problem ~\cite{lim2019enhancing, wood2021trading, wood2022slow, DeepInception}, which have been shown to outperform TSMOM in terms of risk-adjusted returns~\cite{lim2019enhancing}. 

We opt to use factors that are commonly used in trend-following strategies~\cite{AHLMomentum, CenturyOfTrendFollowing, lim2019enhancing}. Concretely, we use returns aggregated and normalized over different time scales:
\begin{equation}
    \hat{r}^{(i)}_{t-t',t} = r^{(i)}_{t-t',t}/\sigma_t^{(i)}\sqrt{t'},
\end{equation}
and include MACD indicators:
\begin{equation}
    \label{eqn:model-inputs}
    \mathbf{x}_t^{(i)} = \mathrm{Concat}\left(\left[\hat{r}^{(i)}_{t-t',t} \, | \, t'\in\{1, 21, 63, 126, 252\} \right], \left[\mathrm{MACD}\left(p_{1:t}^{(i)}, S, L\right) \, | \, \forall \, (S, L) \right]\right).
\end{equation}

The DMN framework simultaneously learns asset price trends and position sizes. The position sizes are estimated using a neural network:
\begin{equation}
    z_{-l_t:t}^{(i)} 
    =  \mathrm{DMN} \left(\mathbf{x}_{-l_t:t}^{(i)}\right)
    =  \left(\tanh \circ ~\mathrm{Linear}  \circ ~\mathrm{g} \right) \left(\mathbf{x}_{-l_t:t}^{(i)}\right),
    \label{eqn:dmn}
\end{equation}
where $g:\mathcal{X}^{l_t}\to\mathbb{R}^{d_h}$ is a neural network followed by a linear mapping, $\mathrm{Linear}: \mathbb{R}^{d_h} \to \mathbb{R}$, and $\tanh$ activation function such that traded position $z_t^{(i)} \in (-1,1)$. The neural network hidden state dimension is denoted $d_h$.

The primary innovation of DMNs, as introduced in~\cite{lim2019enhancing}, was to output traded positions and directly optimize the Sharpe ratio: a risk-adjusted return metric measuring returns per unit volatility. Typically most fund managers or CTAs will have a predefined risk tolerance and aim to maximize returns given this constraint\footnote{The loss function can easily be tailored to instead use drawdown, VaR (Value at Risk), or even some combination of these, instead of volatility as the risk measure.}.
With the aim to optimize neural network parameters $\theta$, the Sharpe loss function is defined as:
\begin{equation}
\mathcal{L}_{\mathrm{Sharpe}}^{\mathrm{DMN}(\cdot)}(\theta) = 
-\sqrt{252}\frac{\mathrm{mean}_\Omega\left[\frac{\sigma_{\mathrm{tgt}}}{\sigma_t^{(i)}}~r_{t+1}^{(i)}~\mathrm{DMN} \left(\mathbf{x}_{-l_t:t}^{(i)}\right)\right]}{\mathrm{std}_\Omega\left[\frac{\sigma_{\mathrm{tgt}}}{\sigma_t^{(i)}}~r_{t+1}^{(i)}~\mathrm{DMN}\left(\mathbf{x}_{-l_t:t}^{(i)}\right)\right]},
\label{eqn:sharpe_loss}
\end{equation}
where $\Omega$ is a batch of $|\Omega|$ pairs $(i,t)\in \mathcal{I}\times \mathcal{T}$. In practice, when training the model, we select $b$ sequences such that $\Omega =  \bigcup \{\{i\} \times ((t-l_t +l_s +1):t) | i \in \mathcal{I}, t \in \mathcal{T}\}$, with warm-up period  $l_s \leq l_t$. That is, our loss function ignores the first $l_s$ predictions for each sampled sequence.

The Long Short-Term Memory cell (LSTM)~\cite{hochreiter1997long} is a popular Recurrent Neural Network (RNN) tailored to modeling sequences and is the primary DMN component~\cite{lim2019enhancing, wood2022slow}. In addition to the output for each time-step $\mathbf{h}^{(i)}_t\in (-1,1)^{d_h}$, the LSTM maintains $\mathbf{c}^{(i)}_t\in\mathbb{R}^{d_h}$, a cell state, which stores longer-term information:
\begin{equation}
    (\mathbf{h}_t^{(i)}, \mathbf{c}_t^{(i)}) = \mathrm{LSTM}(\mathbf{x}_t^{(i)}, \mathbf{h}_{t-1}^{(i)}, \mathbf{c}_{t-1}^{(i)}).
\end{equation}
The LSTM modulates information through a series of gates, clearing $\mathbf{c}^{(i)}_t$ when necessary, such as during regime change, and maintains $\mathbf{h}^{(i)}_t$ as a localized summary of the sequence. The initialization $(\mathbf{h}^{(i)}_0, \mathbf{c}^{(i)}_0)$ can be specific per contract, and we provide details of our implementation~\Cref{sec:seq_representation}. 

\subsection{Attention}
\label{sec:attention}
The attention mechanism computes a weighted average of elements where the weights depend on an input query and the elements’ keys~\cite{vaswani2017attention}. It dynamically decides on which input elements to focus on or ``attend'' to. Specifically, it computes a similarity between the \emph{query} vector and \emph{key} vectors. These vectors can either be model inputs, some deep-learning hidden state, or, in the case of our work, a hidden state summarising a sequence. 
If we have a \emph{query} vector $\mathbf{q} \in \mathbb{R}^{d_q}$ and we want to compute the relative importance of each \emph{key} in $K=\{\mathbf{k}_1,\ldots,\mathbf{k}_{|\mathcal{C}|}\}$, we calculate soft weights with a softmax function:
\begin{equation}
\label{eqn:att_prob}
p_{\mathbf{q}}(\mathbf{k}) = \frac{\alpha(\mathbf{q}, \mathbf{k}) }{\sum_{\mathbf{k}^\prime \in K } \alpha(\mathbf{q}, \mathbf{k}^\prime)}, \quad \alpha(\mathbf{q}, \mathbf{k}) = \exp \left(\frac{1}{\sqrt{d_{\text{att}}}}\langle W_q \mathbf{q} , W_k \mathbf{k} \rangle\right),
\end{equation}
with learnable weight matrices $W_{(\cdot)} \in \mathbb{R}^{d_{\mathrm{att}}\times d_{(\cdot)}}$ and attention dimension $d_{\mathrm{att}}$. The inner-product $\langle \cdot, \cdot \rangle$ is the mechanism used to compute similarity. The primary benefit of attention is that it provides a direct connection with all of the keys and, furthermore, these weights are interpretable. We then use each weight $p_{\mathbf{q}}(\cdot)$ to scale the \emph{values} $V=\{\mathbf{v}_1, \ldots, \mathbf{v}_{|\mathcal{C}|}\}$, where each value has a corresponding key, and aggregate as:
\begin{equation}
    \mathrm{Att}(\mathbf{q}, K, V) = 
    \sum_{j=1}^{|\mathcal{C}|} p_{\mathbf{q}}(\mathbf{k}_j )~W_v~\mathbf{v}_j.
    \label{eqn:attention}
\end{equation}

We elaborate on the specific details of our implementation in~\Cref{sec:x-attn}. We can generalize~\eqref{eqn:attention} to multiple parallel attention heads to make the model bigger, allowing the model to capture representations for different patterns and time-scales. The cross-attention mechanism in X-Trend uses $4$ heads.

\section{Cross-Attentive Time-Series Momentum Forecaster}
\subsection{Sequence Representations and Baseline Neural Forecaster}
\label{sec:seq_representation}
We want to create sequence summaries, of sequence length $l$, with a learnable function $\Xi:\mathcal{X}^l \times \mathcal{S} \to \mathbb{R}^{d_h}$, where $d_h$ is our hidden dimension. Our representation is created to summarise not just input sequences $\mathbf{x}_{-l_t:t}^{(i)}\in\mathcal{X}^l$, but also side information $s^{(i)}\in\mathcal{S}$, namely the category (ticker) of the futures contracts. We include this category because we know the dynamics of different contracts can vary significantly; for example crude oil compared to the 5-year treasury bond. For each time-step, and each asset, we input features  $\mathbf{x}^{(i)}_t\in\mathcal{X}$, as defined in~\cref{eqn:model-inputs}. 

We encode side information $s^{(i)}$, with an entity embedding: a learnable mapping of each category into $\mathbb{R}^{d_h}$, which can automatically learn to group similar assets in the embedding space~\cite{EntityEmbeddings}. We use a feedforward (FFN) network to fuse the entity embeddings and time-series representations. Throughout our paper we use $\mathrm{ELU}$ (Exponential Linear Unit)~\cite{clevert2015fast} activations which are continuously differentiable everywhere and avoid dead neurons (which are completely deactivated). We indicate the ability to \textcolor{blue}{optionally include static information} via embeddings in \textcolor{blue}{blue}:
\begin{subequations}
\begin{align}
    \mathrm{FFN} (\mathbf{h}_t\textcolor{blue}{, s}) &= \mathrm{Linear}_{3} \circ \mathrm{ELU}\left(\mathrm{Linear}_{1}(\mathbf{h}_t) \textcolor{blue}{+ \mathrm{Linear}_{2}(\mathrm{Embedding}(s)}\right),
\end{align}
\end{subequations}
where $\mathrm{Linear}_{(\cdot)}(\cdot)$ is a learnable linear transformation into $\mathbb{R}^{d_h}$ and $\mathbf{h}_t$ is a vector which can either be a hidden-state or a vector. Additionally, we use the Variable Selection Network (VSN)~\cite{lim2019enhancing}, which weights out different lagged normalized returns and MACD features. 

We associate a learnable nonlinear function $\mathrm{FFN}_{j}: \mathbb{R} \rightarrow \mathbb{R}^{d_h}$ with the $j$-th element of $\mathbf{x}_{t}$, and scale by the associated learnable weight $w_{t,j}$, of $\mathbf{w}_t \in \mathbb{R}^{|\mathcal{X}|}$:
\begin{subequations}
\begin{align}
    \mathbf{w}_t &=  \mathrm{Softmax}\circ \mathrm{FFN}(\mathbf{x}_t,s) \\
     \mathrm{VSN}(\mathbf{x}_t)&=\sum_{j=1}^{|\mathcal{X}|} w_{t,j}~\mathrm{FFN}_{j}(x_{t, j}),
\end{align}
\end{subequations}
where the $j$-th element of the softmax is defined as: $\mathrm{Softmax}(\mathbf{x})_j = e^{x_j} / \sum_{k=1}^{|\mathcal{X}|} e^{x_k}$.

Our model consists both of an encoder and a decoder. It is important to note that in the decoder we also include the output of the encoder, $\mathbf{y}^{(i)}_{-l_t:t}\in \mathcal{Y}^{l_t}$, as an input, where  $\mathcal{Y} \subseteq \mathbb{R}^{d_h}$ (\Cref{sec:loss-function}). Our encoder sequence representations, $\Xi(\cdot,\cdot)$, can be summarised as:
\begin{subequations}
\label{eqn:seq-rep}
\begin{align}
    \mathbf{x}_t' &= \mathrm{VSN}(\mathbf{x}_t, s), \label{eqn:vsn_step_and_decoder}\\
    (\mathbf{h}_t, \mathbf{c}_t) &= \mathrm{LSTM}(\mathbf{x}_t', \mathbf{h}_{t-1}, \mathbf{c}_{t-1}), \\
    \mathbf{a}_t &= \mathrm{LayerNorm}(\mathbf{h}_t + \mathbf{x}_t') \label{eqn:lstm-skip-connection}\\
    \Xi(\mathbf{x}_{-l:t}, s)  &= \mathrm{LayerNorm} \left( \mathrm{FFN}_2 (\mathbf{a}_t, s) + \mathbf{a}_t\right). \label{eqn:ffn-skip-connection}
\end{align}
\end{subequations} 
The LSTM initial state is learnable and specific to each contract, setting $(\mathbf{h}_0, \mathbf{c}_0) = (\mathrm{FFN}_3 \circ \mathrm{Embedding}(s), \mathrm{FFN}_4 \circ \mathrm{Embedding}(s))$. The skip connections, in~\eqref{eqn:lstm-skip-connection} and~\eqref{eqn:ffn-skip-connection}, allow for the respective components to be suppressed, enabling the model to be as complex as necessary. See~\cref{fig:x-attentive-encoder} for an overview of our model.

The \emph{Baseline} neural forecaster which we compare to uses $\Xi(\cdot, \cdot)$ as a model i.e. $g(\cdot)$ from \eqref{eqn:dmn}. The baseline architecture consists of only the decoder. The X-trend model adds an encoder and the cross-attentive steps. 

\subsection{Context Set Construction}
\label{sec:context_set_construction}

\begin{figure}[h]
    \centering
    \includegraphics[width=0.9\textwidth]{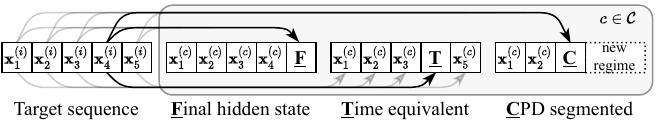}
    \caption{An illustration of the different ways the target is able to attend to the context set. \textbf{\underline{F}}, every hidden state in the target sequence is able to attend to the final hidden states of the contexts. \textbf{\underline{T}}, the time equivalent hidden state in the target is able to attend to the corresponding hidden state in the contexts. \textbf{\underline{C}}, every hidden state in the target sequence is able to attend to the final hidden state in the change-point segmented contexts. The dark arrows illustrate the context time-steps the $4$-th target time-step attends to.}
    
    \label{fig:context-hidden}
\end{figure}

\begin{figure}[h]
    \centering
    \includegraphics[width=0.9\textwidth]{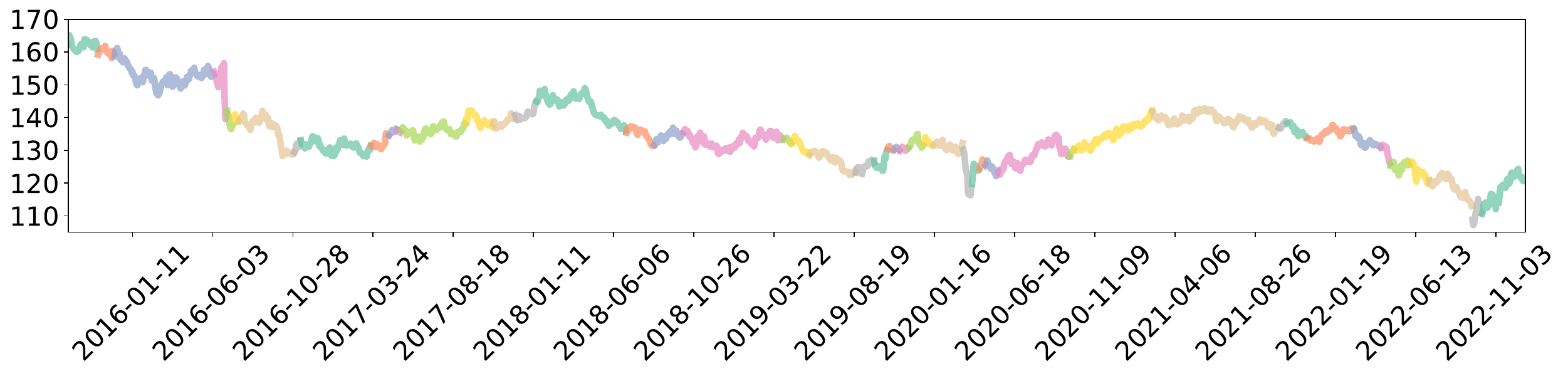}
    \caption{A time-series segmented with change-point detection to create sequences for the context set. Different colours are different regimes. This example shows the British Pound Sterling continuous, ratio-adjusted, futures contract. Here, for illustrative purposes, regimes are segmented with a change-point threshold of $L_C / (L_M + L_C) \geq 0.99$, where $L_M$ is the likelihood of fitting a Gaussian Process characterized by a Matérn 3/2 kernel, and $L_C$ is another characterized by a Change-point kernel. Details of this procedure can be found in~\Cref{app:cpd}.}
    \label{fig:cpd-segmentation}
\end{figure}

When backtesting, we randomly sample a context set of size $|\mathcal{C}|$ sequences from the past\footnotemark{} to enforce causality at the time of prediction for our target problem. We explore three different approaches to constructing our context sequences and choosing the point $t_c$ at which we condition on, illustrating these in~\cref{fig:context-hidden}:
\footnotetext{During training we do not enforce causality when we construct our context set, as the aim is to teach the model to best transfer patterns; however, the training set only contains information prior to the test date to ensure our model is causal at test time.} 
\begin{enumerate}
    \item \textbf{Final hidden state and random sequences of fixed length.} We sample random sequences across time and assets of fixed length $l_c$ and we condition on the final hidden states of the sequences, which summarises the sequence. 
    \item \textbf{Time-equivalent hidden state.} We sample random context sequences that are the same length as the target sequence: $l_c = l_t$. For each time-step in the target sequence, we condition on the time-equivalent hidden states i.e. the $k$-th target step attends to the $k$-th context steps. This allows the model to incorporate additional adjacent regimes by conditioning on different representations at each time-step of the target sequence.
    \item \textbf{Change-point detection (CPD) segmented representations.} We use a Gaussian Process change-point detection algorithm, detailed in~\Cref{app:cpd}, to segment the context set into regimes. An example of this segmentation for the British Pound can be seen in~\cref{fig:cpd-segmentation}. We randomly sample $|\mathcal{C}|$ change-point segmented sequences and condition on the final hidden states of these change-point time-series segments. We limit to a maximum sequence size. We test both $21$ day ($1$ month) and $63$ day ($3$ month) maximum length sizes, using a higher CPD severity threshold for the $63$ day version, calibrating a change-point threshold
    in both cases such that the average sequence length is approximately half of the maximum length.
\end{enumerate}

\subsection{Cross Attention}
\label{sec:x-attn}
In order for our predictions to leverage a context set we use a cross-attention mechanism between our target sequence and a context set of sequences. This allows our target to attend to different sequences from different assets across time and across different regimes. The rationale is that this context set contains a much broader range of patterns in comparison to the near-term history of the same asset, which was successfully leveraged with attention by~\cite{wood2021trading} (\Cref{sec:deep-learning-ts-related}).  

\begin{figure}[htb]
    \centering
    \includegraphics[width=0.9\textwidth]{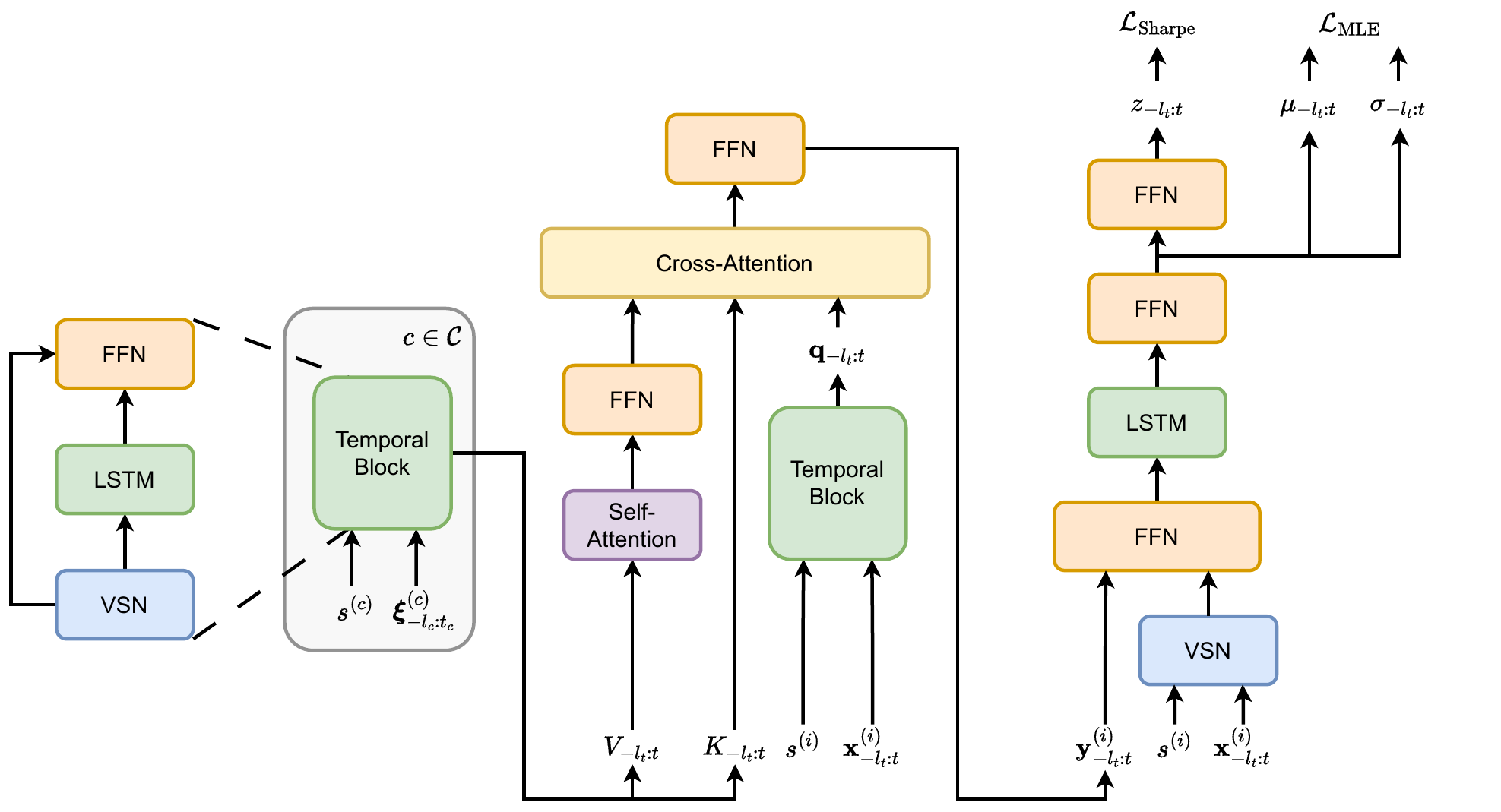}
    \caption{Encoder and decoder X-Trend-G model.  The FFN, VSN, Self-Attention and Cross-Attention components are all applied element-wise to each time-step. Sequences in the context set are mapped to representations via $\Xi_{\text{key}}(\cdot, \cdot)$ and $\Xi_{\text{value}}(\cdot, \cdot)$. For the key inputs we exclude next-day returns and use $\mathbf{x}_{-l_c:t_c}^{(c)}$ instead of $\boldsymbol{\xi}_{-l_c:t_c}^{(c)}$. Contexts are then passed to the cross-attention as keys and values with a representation of the target sequence $\mathbf{x}_{t'}^{(i)}$ which we want to make forecasts as the query. It should be noted that we have a separate instance of keys $K_{t'}$ and values $V_{t'}$ for the query $q_{t'}$ at each time-step $t'\in (t-l_t+1:t)$,  which we detail in~\cref{fig:context-hidden}.  The decoder then takes the target sequence and the output representation from the encoder, $\mathbf{y}_{t'}^{(i)}$. It outputs a position for the Sharpe stream and the forecast stream, which we label $(\mu_{t'}, \sigma_{t'})$, for the maximum likelihood. Side information $s^{(i)}$ regarding the target asset is also passed as input to the decoder for few-shot learning only, not zero-shot learning. If we are not using the joint loss function, we instead output for the Sharpe stream after the second last FFN.  
    }
    \label{fig:x-attentive-encoder} 
\end{figure}

We create the set of key vectors, $K_{t}$, and set of value vectors, $V_{t}$:
\begin{equation}
    K_{t} = \left\{\Xi_{\text{key}}(\mathbf{x}^{(c)}_{-l_c:t_c},s^{(c)})\right\}_{c=1}^{|\mathcal{C}|}, \quad
    V_{t} = 
    \left\{\Xi_{\text{value}}(\bm{\xi}^{(c)}_{-l_c:t_c},s^{(c)})\right\}_{c=1}^{|\mathcal{C}|},
\end{equation}
where $\mathbf{x}^{(c)}_{-l_c:t_c}$ denotes context sequence without the target returns. For a given query, we calculate attention weights for each key (see \cref{eqn:att_prob}). Then adapting~\cref{eqn:attention}, we weight the value vectors by the attention weights and expand to multiple concurrent heads to increase the representational space. We leverage the context set via the following steps: 
\begin{align}
	\mathbf{q}_t &= \Xi_{\mathrm{query}}(\mathbf{x}_{-l_t:t}^{(i)}, s^{(i)}) & \text{(query representation)} \\
	V_{t}' &= \mathrm{FFN}_{1} \circ \mathrm{Att}_{1}(V_{t},V_{t},V_{t}) & \text{(self-attention)} \\
    \mathbf{y}^{(i)}_t &= \mathrm{LayerNorm} \circ \mathrm{FFN}_{2} \circ \mathrm{Att}_{2}(\mathbf{q}_t, K_{t}, V_{t}'). & \text{(cross-attention)}
\end{align}
We illustrate this in~\cref{fig:x-attentive-encoder}. The self-attention step, which outputs the updated set of values $V_{t}'$,  helps to identify similarities between regimes within the context set.

\subsection{Decoder and Loss Function}
\label{sec:loss-function}
Similarly to our sequence representations in the encoder, \cref{eqn:seq-rep}, we summarize our target sequence in the decoder. This time our sequence representation, $\Xi_{\textcolor{magenta}{\mathrm{Dec}}}: \mathcal{X}^{l_t} \times \mathcal{S} ~\textcolor{magenta}{ \times ~ \mathcal{C}} \to \mathbb{R}^{d_h}$, fuses the output of the encoder, $\mathbf{y}^{(i)}_{-l_t:t}$. Highlighting the additional components in \textcolor{magenta}{magenta}, our \textcolor{magenta}{decoder} representation is computed as:
\begin{subequations}
\label{eqn:seq-refs}
\begin{align}
        \mathbf{x}_t' &= \textcolor{magenta}{\mathrm{LayerNorm} \circ \mathrm{FFN}_1 \circ \mathrm{Concat}(}\mathrm{VSN}(\mathbf{x}^{(i)}_t, s^{(i)})\textcolor{magenta}{, \mathbf{y}^{(i)}_t)}, \label{eqn:vsn_step_and_decoder}\\
    (\mathbf{h}_t, \mathbf{c}_t) &= \mathrm{LSTM}(\mathbf{x}_t', \mathbf{h}_{t-1}, \mathbf{c}_{t-1}), \\
    \mathbf{a}_t &= \mathrm{LayerNorm}(\mathbf{h}_t + \mathbf{x}_t') \label{eqn:lstm-skip-connection}\\
    \Xi_{\textcolor{magenta}{\mathrm{Dec}}}(\mathbf{x}^{(i)}_{-l_t:t}, s^{(i)} \textcolor{magenta}{, \mathbf{y}^{(i)}_{-l_t:t}})  &= \mathrm{LayerNorm} \left( \mathrm{FFN}_2 (\mathbf{a}_t, s^{(i)}) + \mathbf{a}_t\right). \label{eqn:ffn-skip-connection}
\end{align}
\end{subequations} 

We want to model the next-day volatility scaled return. This is a regression task where we parameterize the predictive mean via $\mu: \mathcal{X}^{l_t} \times \mathcal{S} \times \mathcal{C} \to \mathbb{R}$ and volatility via $\sigma: \mathcal{X}^{l_t} \times \mathcal{S} \times \mathcal{C} \to \mathbb{R}^+$; the likelihood is
$p \left (\frac{\sigma_{\mathrm{tgt}}}{\sigma_t^{(i)}}r_{t+1}^{(i)} \mid \mathbf{x}^{(i)}_{-l_t:t}, s^{(i)}, \mathcal{C} \right ) = \mathcal{N}\left(\frac{\sigma_{\mathrm{tgt}}}{\sigma_t^{(i)}}~r_{t+1}^{(i)} ;\mu(\mathbf{x}^{(i)}_{-l_t:t}, s^{(i)}, \mathcal{C}), \sigma(\mathbf{x}^{(i)}_{-l_t:t}, s^{(i)}, \mathcal{C})\right)$. Thus our loss function is to minimize the log-likelihood under a Gaussian distribution of future returns:
\begin{align}
\mathcal{L}_{\mathrm{MLE}}(\theta) &= -\frac{1}{|\Omega|}\sum_{(t,i)\in\Omega} \log p\left(\frac{\sigma_{\mathrm{tgt}}}{\sigma_t^{(i)}}~r_{t+1}^{(i)} | \mathbf{x}^{(i)}_{-l_t:t}, s^{(i)}, \mathcal{C} \right)
\end{align}

It is important to note that the Gaussianity of returns in financial time-series is a convenient approximation, which in practice does not always hold, especially in the tail. However, since we are focused on optimizing the Sharpe ratio in this paper\footnotemark{}, predictive mean and volatility estimates of next-day return outputs are very useful since Sharpe is dependent on realized returns and volatility. 

\footnotetext{Due to the non-Gaussianity of returns observed in practice, the Sharpe ratio often may not be the best metric for measuring risk-adjusted returns.}
Rather than directly optimizing Sharpe with a DMN, we propose to jointly optimize the likelihood for next-day predictions and the Sharpe. The Sharpe loss requires an additional neural network head, \textbf{P}redictive distribution (mean and standard deviation) \textbf{T}o \textbf{P}osition: head $\mathrm{PTP}_{\mathrm{G}}:\mathbb{R}^2\to(-1,1)$. This is a single FFN followed by a $\tanh$ activation. Our joint loss function is:
\begin{equation}
    \label{eqn:joint-gauss-loss}
    \mathcal{L}_{\mathrm{Joint}}^{\mathrm{MLE}}(\theta) = \alpha ~\mathcal{L}_{\mathrm{MLE}}(\theta)  +\mathcal{L}_{\mathrm{Sharpe}}^{\mathrm{PTP_G}(\cdot)}(\theta),
\end{equation}
where $\mathcal{L}_{\mathrm{Sharpe}}^{\mathrm{PTP_G}(\cdot)}(\theta)$ is~\cref{eqn:sharpe_loss} applied to the PTP outputs and $\alpha$ is a tunable hyperparameter to balance the two loss functions.

As an alternative to assuming Gaussianity of returns, we can instead perform Quantile Regression (QRE). The aim of QRE is to learn the full probability distribution of next-day returns. QRE has proven successful for multi-step time-series regression~\cite{wen2017multi}. For quantiles $\eta \in \mathcal{H}$, pairs $(t,i)\in\Omega$  and target $\tilde{r}^{(i)}_{t+1}=\frac{\sigma_{\mathrm{tgt}}}{\sigma_t^{(i)}}r_{t+1}^{(i)}$, our QRE loss function is:
\begin{equation}
\mathcal{L}_{\text{QRE}}(\theta) = \frac{1}
{|\Omega \times \mathcal{H}|}
\sum_{\Omega \times \mathcal{H}}
\left[ \eta ~\left(\tilde{r}^{(i)}_{t+1} - Q_{\eta}(\mathbf{x}^{(i)}_{-l_t:t}, s^{(i)}, \mathcal{C})\right)_+ 
+ \left(1 - \eta\right) \left(Q_{\eta}(\mathbf{x}^{(i)}_{-l_t:t}, s^{(i)}, \mathcal{C}) - \tilde{r}^{(i)}_{t+1}\right)_+ \right],
\end{equation}
where $(\cdot)_{+}=\max (0, \cdot)$ and $Q:\mathcal{X}^{l_t}\times \mathcal{S} \times \mathcal{C} \to\mathbb{R}^{|\mathcal{H}|}$ is a neural network head which we replace $\mu(\cdot, \cdot, \cdot)$ and $\sigma(\cdot, \cdot, \cdot)$ with. Our set of quantiles, $\mathcal{H} = \{0.01, 0.05, 0.1, 0.2, 0.3, 0.4, 0.5, 0.6, 0.7, 0.8, 0.9, 0.95, 0.99\}$, includes quantiles in the left and right tail -- the market movements which typically have the largest impact on our strategy risk-adjusted returns. We define the joint QRE loss function, $\mathcal{L}_{\mathrm{Joint}}^{\mathrm{QRE}}(\cdot)$ as:
\begin{align}
    \label{eqn:joint-qre-loss}
    \mathcal{L}_{\mathrm{Joint}}^{\mathrm{QRE}}(\theta) = \alpha ~\mathcal{L}_{\mathrm{QRE}}(\theta)  +\mathcal{L}_{\mathrm{Sharpe}}^{\mathrm{PTP_{Q}}(\cdot)}(\theta).
\end{align}
with the PTP for the Sharpe loss again a FFN, $\mathrm{PTP}_{\mathrm{Q}}:\mathbb{R}^{|\mathcal{H}|}\to(-1,1)$.

Utilizing the outputs of our encoder, $\mathbf{y}^{(i)}_{-l_t:t}$, our decoder outputs a trading signal as follows:
\begin{equation}
     z_t = \mathrm{PTP} \circ \Xi_{\mathrm{Dec}}(\mathbf{x}^{(i)}_{-l_t:t}, s^{(i)}, \mathbf{y}^{(i)}_{-l_t:t}).
\end{equation}
We refer to the \textbf{G}aussian MLE (Maximum Likelihood Estimation) variant of the architecture as X-Trend-\textbf{G}, the \textbf{Q}RE variant as X-Trend-\textbf{Q} and the Sharpe loss variant as X-Trend.

It is important to note that in the \emph{zero-shot} setting we exclude the ticker-type embedding of $s^{(i)}$ for the target sequence. This is because we are trading a previously unseen asset and we have not yet seen any contract-specific dynamics. We do however still include this information in the context set, where the aim is to quickly identify similarities with previously seen contracts.

\section{Experiments}
The cross-attention mechanism is key to obtaining low error forecasts for few-shot learning for a toy dataset of Gaussian Process (GP) draws~\cite{garnelo2018neural}. We implemented the recurrent attentive neural process~\cite{qin2019recurrent} for this toy dataset, and we ablate certain components and we show that the cross-attention mechanism is key to obtaining low-error few-shot forecasts. This result motivates using a cross-attention for momentum. See~\Cref{sec:gp_draws_exp} for further details. 

We backtest our X-Trend variants on a portfolio of $50$ of the most liquid, continuous futures contracts\footnote{The futures contracts are chained together using the backwards ratio-adjusted method.}, $\mathcal{I}$, over the period from 1990 to 2023, extracted from the Pinnacle Data Corp CLC Database\footnote{\url{https://pinnacledata2.com/clc.html}}. The futures contracts we have selected are amongst the most liquid and typical for backtesting TSMOM strategies~\cite{moskowitz2012time, CenturyOfTrendFollowing, lim2019enhancing, wood2021trading}. To test-out-of-sample across the entire history, we use an expanding window approach, where we initially train on $1990$ to $1995$, test out-of-sample on the period from $1995$ to $2000$, expand the training window to from $1990$ to $2000$, then test out-of-sample on the subsequent $5$ years and so on. We take particular note of performance over the $2020$ to $2022$ period, which covers the COVID-19 crisis, exhibiting dynamics that are significantly different to the training set.

For our zero-shot experiments, we randomly selected $20$ of the $50$ Pinnacle dataset assets as the test set, $\mathcal{I}_{ts}$, leaving the other $30$, $\mathcal{I}_{tr}$ for the context set and training. Despite the fact these futures contracts had historical data available in reality, we constructed this experiment as an artificial zero-shot setting to provide insight into transferability to unseen assets. Furthermore, this low-resource setting is particularly challenging because $\mathcal{I}_{tr}$ is also small at only $30$ contracts. The dataset is described in further detail in~\Cref{app:data}. The details of training the neural networks are provided in~\Cref{app:expt-details}. 

\section{Results}
\label{sec:results}
\begin{figure}[hb]
    \centering
    \includegraphics[width=\textwidth]{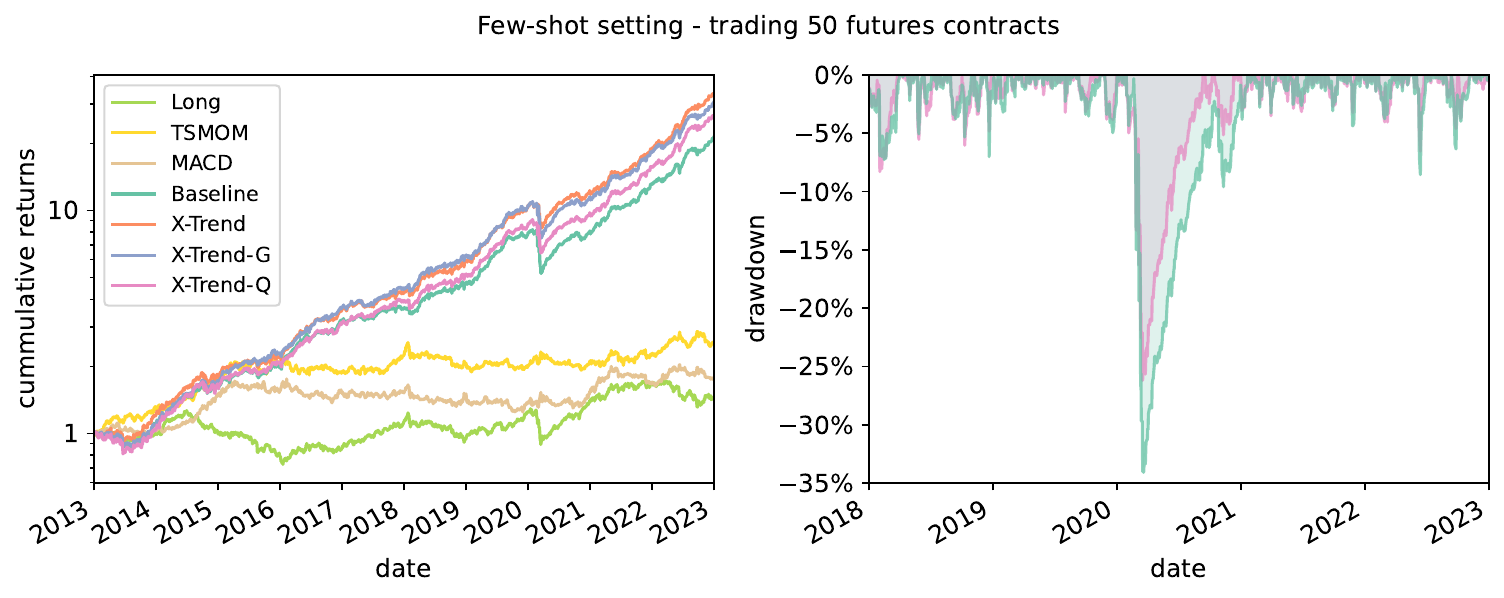}
    \caption{Few-shot setting cumulative strategy returns (left) and drawdown plot (right), averaged across 10 full repeats and an additional portfolio volatility re-scaling step to 15\% volatility. We only plot drawdown of the Base Learner and X-Trend-Q, the primary comparison, to reduce clutter.}
    \label{fig:few-shot-returns}
\end{figure}
\begin{figure}[hb!]
    \centering
    \includegraphics[width=\textwidth]{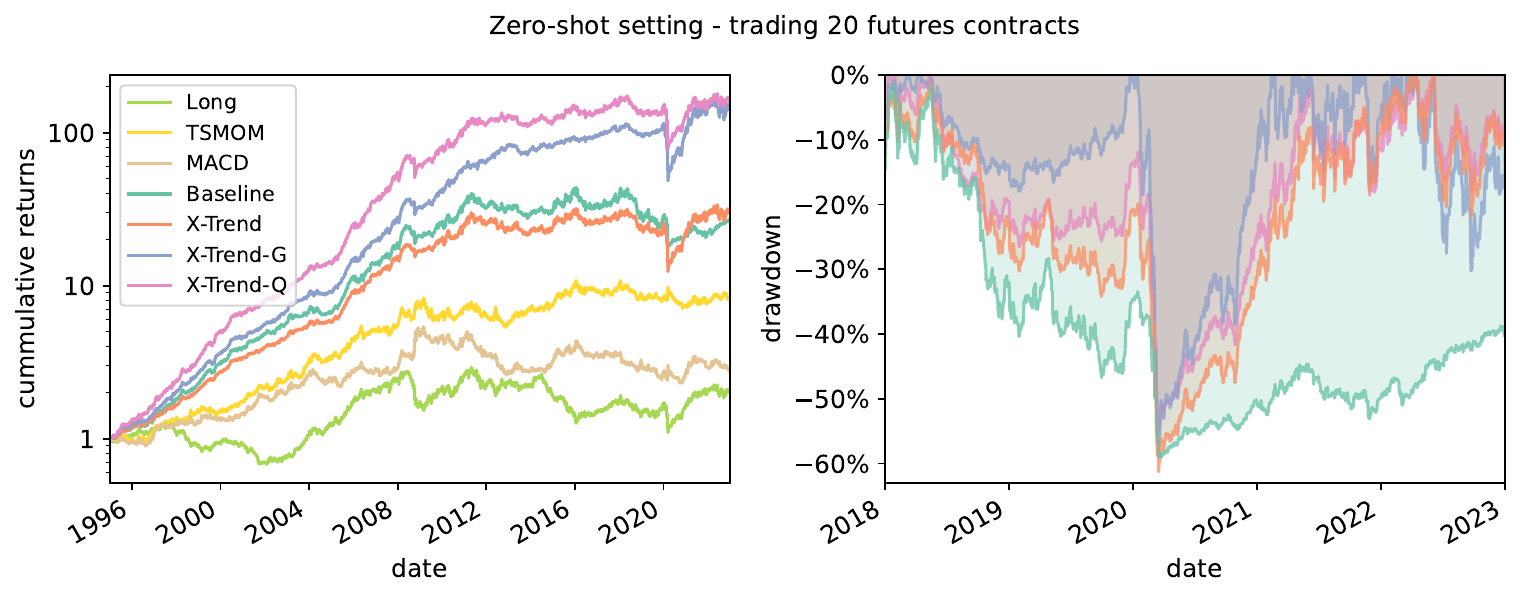}
    \caption{Zero-shot setting cumulative strategy returns (left) and drawdown plot (right), averaged across 10 full repeats and an additional portfolio volatility re-scaling step to 15\% volatility.}
    \label{fig:zero-shot-returns}
\end{figure}

\begin{table}[t]
\centering
\small
\begin{tabular}{lcccccccc}
\toprule
& & \textbf{Context Time-step} &  & & \multicolumn{3}{c}{\textbf{Average Annual Sharpe Ratio}} & \\
\textbf{Method} & \textbf{Loss}
& \textbf{\underline{F}inal/\underline{T}ime/\underline{C}PD} & $|\mathcal{C}|$ &  $l_c$
& \textbf{$2018$--$2023$} & \textbf{$2013$--$2023$} & \textbf{$1995$--$2023$} \\
\hline
\multirow{3}{*}{\textbf{Reference}} & Long & & & &  0.48 & 0.40 & 0.60 \\
 & TSMOM & & & &  0.23 & 0.71 & 1.01 \\
 & MACD & & & &  0.27 & 0.45  & 0.71  \\
\hline
\multirow{3}{*}{\textbf{Baseline}} & Sharpe & & & & \textbf{2.27} & \textbf{1.93} & \textbf{2.91} \\
 & J-Gauss & & & & \textbf{2.43 (+7.2\%)} & \textbf{2.06 (+7.0\%)}& \textbf{3.04 (+4.5\%)} \\
 & J-QRE & & & & 2.26 (-0.5\%) & 1.96 (+1.6\%) & 2.89 (-0.9\%) \\
\hline
\multirow{9}{*}{\textbf{X-Trend}} & Sharpe & T & 10 & 126 & 2.28 (+0.2\%) & 1.97 (+2.4\%) & 2.93 (+0.5\%) \\
& Sharpe & F & 10 & 21 & 2.35 (+3.4\%) & 1.99 (+3.2\%) & 3.01 (+3.4\%)\\
& Sharpe & F & 20 & 21 & 2.38 (+4.9\%) & 1.99 (+3.4\%) & 3.02 (+3.6\%)\\
& Sharpe & F & 30 & 21 & 2.25 (-0.9\%) & 1.99 (+3.1\%) & 3.03 (+4.0\%)\\
& Sharpe & F & 10 & 63 & 2.31 (+1.7\%) & 1.97 (+2.3\%) & 3.11 (+6.9\%)\\
& Sharpe & C & 10 & 21 & 2.30 (+1.1\%) & 2.02 (+4.7\%) & 3.04 (+4.5\%)\\
& Sharpe & C & 20 & 21 & \textbf{2.65 (+16.9\%)} & \textbf{2.17 (+12.5\%)} & \textbf{3.17 (+8.8\%)}\\
& Sharpe & C & 30 & 21 & 2.38 (+5.0\%) & 2.02 (+4.6\%) & 3.08 (+5.8\%)\\
& Sharpe & C & 10 & 63\textsuperscript{**} & 2.50 (+10.1\%) & 2.14 (+11.2\%) & 3.11 (+6.8\%)\\
\hline
\multirow{7}{*}{\textbf{X-Trend-G}} & J-Gauss & T & 10 & 126 & 2.47 (+8.8\%)& \textbf{2.14 (+11.1\%)}& 3.16 (+8.5\%)  \\
& J-Gauss & F & 10 & 21 & 2.25 (-1.0\%)& 1.90 (-1.3\%)& 3.10 (+6.5\%)  \\
& J-Gauss & F & 20 & 21 & \textbf{2.52 (+11.0\%)}& 2.10 (+8.8\%)& 3.05 (+4.8\%)  \\
& J-Gauss & F & 30 & 21 & 2.26 (-0.5\%)& 2.08 (+7.9\%)& 3.04 (+4.2\%)  \\
& J-Gauss & F & 10 & 63 & 2.42 (+6.5\%)& 2.07 (+7.4\%)& 3.17 (+8.8\%)  \\
& J-Gauss & C & 10 & 21 & 2.42 (+6.8\%)& 2.09 (+8.1\%)& 3.06 (+5.1\%)  \\
& J-Gauss & C & 20 & 21 & 2.42 (+6.5\%)& 2.03 (+5.2\%)& 3.11 (+6.8\%)  \\
& J-Gauss & C & 30 & 21 & 2.51 (+10.4\%)& 2.12 (+9.8\%)& 3.07 (+5.4\%)  \\
& J-Gauss & C & 10 & 63\textsuperscript{**}& 2.32 (+2.1\%) & 2.00 (+3.6\%) &\textbf{ 3.18 (+9.2\%)} \\
\hline
\multirow{7}{*}{\textbf{X-Trend-Q}} & J-QRE & T & 10 & 126 &	2.53 (+11.6\%) & 	2.12 (+10.1\%) & 	3.08 (+5.9\%) \\
& J-QRE & F & 10 & 21 & 	2.38 (+4.9\%) & 	1.98 (+2.5\%) & 	3.07 (+5.3\%) \\ 
& J-QRE & F & 20 & 21 & 	2.21 (-2.8\%) & 	1.86 (-3.4\%) & 	2.94 (+0.8\%) \\ 
& J-QRE & F & 30 & 21 &	2.42 (+6.6\%) & 	2.05 (+6.5\%) & 	3.03 (+4.2\%) \\ 
& J-QRE & F & 10 & 63 & 	2.49 (+9.6\%) & 	2.04 (+5.8\%) & 	3.06 (+5.1\%) \\ 
& J-QRE & C & 10 & 21 &  	2.26 (-0.5\%) & 	2.03 (+5.3\%) & 	3.07 (+5.3\%) \\ 
& J-QRE & C & 20 & 21 &  	2.53 (+11.6\%) & 	2.08 (+7.7\%) & 	3.07 (+5.5\%) \\ 
& J-QRE & C & 30 & 21 & 	2.4 (+5.6\%) & 	2.02 (+4.5\%) & 	3.01 (+3.2\%) \\ 
& J-QRE & C & 10 & 63\textsuperscript{**} &  	\textbf{2.70 (+18.9\%)} & 	\textbf{2.14 (+10.9\%)} &	\textbf{3.11 (+6.8\%)} \\
\bottomrule
\end{tabular}

\caption{Few-shot learning strategy results, backtested out-of-sample on a portfolio of 50 liquid continuous futures contracts, averaged over ten seeds. We provide an ablation of the X-Trend architectural innovations and provide results for different hyperparameters. It should be noted that $l_c$ is the maximum context length for the CPD segmented version. \textsuperscript{**}We use a higher CPD threshold, such that our context `regimes' are longer.}
\label{tab:results-few-shot}
\end{table}
\begin{table}[t!]
\centering
\small
\begin{tabular}{lcccccccc}
\toprule
& & \textbf{Context Time-step} &  & & \multicolumn{3}{c}{\textbf{Average Annual Sharpe Ratio}} & \\
\textbf{Method} & \textbf{Loss}
& \textbf{\underline{F}inal/\underline{T}ime/\underline{C}PD} & $|\mathcal{C}|$ &  $l_c$
& \textbf{$2018$--$2023$} & \textbf{$2013$--$2023$} & \textbf{$1995$--$2023$} \\
\hline
\multirow{3}{*}{\textbf{Reference}} & Long & & & & 0.28 & 0.02 & 0.28 \\
 & TSMOM & & & &  -0.26 & 0.05 & 0.61 \\
 & MACD & & & &  -0.14 & 0.11 & 0.32 \\
\hline
\multirow{3}{*}{\textbf{Baseline}} & Sharpe & & & & -0.11 & 0.02 & 1.00 \\
 & J-Gauss & & & &  $0.14$ & $0.19$ & $1.25$ \\
 & J-QRE & & & &  $0.16$ & $0.10$ & $1.26$ \\
\hline
\multirow{3}{*}{\textbf{X-Trend}} & Sharpe & C & 20 & 21 & $0.13$ & $0.18$ & $1.17$ \\
 & J-Gauss & C & 20 & 21 & $\mathbf{0.47}$ & $\mathbf{0.47}$ & $\mathbf{1.44}$ \\

 & J-QRE & C & 10 & 63\textsuperscript{**} &  $0.12$ & $0.18$ & $1.27$ \\
\bottomrule
\end{tabular}

\caption{Zero-shot learning strategy results, backtested on a portfolio of 20 liquid continuous futures contracts, averaged over ten seeds. We provide an ablation of the X-Trend architectural innovations and provide results for different hyperparameters.  It should be noted that $l_c$ is the maximum context length for the CPD segmented version. \textsuperscript{**}We use a higher CPD threshold, such that our context `regimes' are longer.}
\label{tab:results_zero_shot}
\end{table}

\subsection{Few-shot Setting}
We plot the few-shot strategy returns over the past $10$ years ($2013$ to $2023$) in~\cref{fig:few-shot-returns}, which is where we start to see larger gains when backtesting our X-Trend strategy, in comparison to the baseline. In particular, we draw attention to the results over the past five years ($2018$ to $2023$), which is an extremely interesting period for few-shot learning because it exhibits significant market turbulence, previously unseen market dynamics, and numerous regime shifts. It includes the Bull market of $2018$/$2019$, followed by the COVID-19 pandemic in $2020$/$2021$ and then the beginning of the Russia-Ukraine war in $2022$. During this $5$-year period X-Trend improves upon the Sharpe of the baseline strategy by $16.9$\% and X-Trend-Q improves upon the baseline by $18.9$\%. It is evident that the cross-attention step is the primary driver of the improvement in risk-adjusted returns. We observe that X-Trend-Q and X-Trend outperform X-Trend-G (\cref{tab:results-few-shot}), which suggests that the model is able to learn and benefit from a more complex returns distribution, compared to assuming Gaussian returns. We argue that this is likely because we explicitly force the model to pay attention to large movements by performing QRE on quantiles in the left (and right) tail of the returns distribution.

The few-shot results in~\cref{tab:results-few-shot} show the impact of changing the context set size, context sequence (maximum) length, and different methodologies for selecting the context hidden state to attend to. Notably, for X-Trend, we demonstrate that by segmenting the context set with CPD, we improve Sharpe by a further $11.3$\%, showing the benefit of constructing context sets with regimes. In effect, we are constructing a context set with only the most informative observations.  

In~\cref{fig:few-shot-returns} we plot the strategy drawdown of the X-Trend-Q strategy compared to our baseline learner over the period $2018$ to $2023$, focusing on the COVID-19 drawdown -- the most significant ``momentum crash'' or drawdown in the TSMOM strategies over the entire $33$ years of historical data. Noting that the drawdown begins on $29$ Jan 2020 for both strategies, we observe a maximum drawdown (MDD) of $26$\% for X-Trend-Q compared to an MDD of $34$\% for the baseline. Furthermore, the drawdown ends after $162$ trading days for X-Trend-Q compared to $254$ days for the Baseline which is an entire year and almost twice as long as X-Trend-Q. This quick recovery demonstrates the ability of our agent to adapt to new regimes.

\subsection{Zero-shot Setting}
For our zero-shot experiments in~\cref{tab:results_zero_shot}. The strategy improvement over the baseline is more pronounced than in the few-shot setting. We plot the strategy returns over the entire backtest ($1995$ to $2023$) in~\cref{fig:zero-shot-returns}, to demonstrate viability across time. We note a degradation of performance $2020$ to $2023$ and, again, we plot drawdowns over the period $2018$ to $2023$, to draw attention to the COVID-19 ``momentum crash''. Despite the fact that the strategy degrades, trading unseen assets during this period is an extremely challenging setting and not only is the strategy still profitable, it outperforms all benchmarks.

Unlike the few-shot setting, it is evident that the joint loss function is the key driver of the improvement of results in the zero-shot setting. Furthermore, in the zero-shot setting, X-Trend-G outperforms X-Trend-Q, indicating that the simpler assumption of Gaussian returns is favourable in a low-resource setting.

\clearpage
\subsection{Dissecting the X-Trend Architecture}
\begin{figure}[b!]
    \centering
    \includegraphics[width=0.9\textwidth]{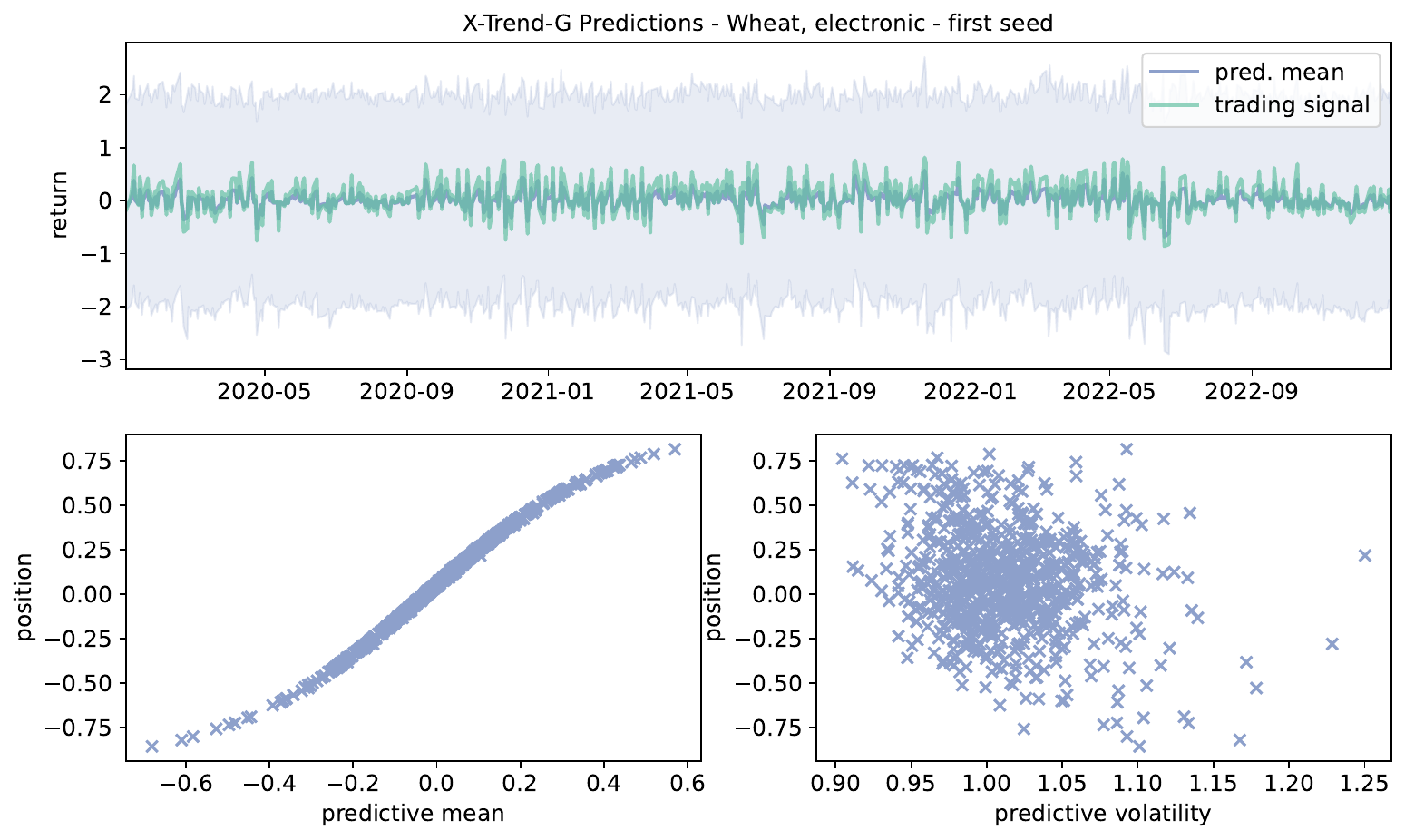}
    \caption{The relationship between X-Trend-G predictive mean and volatility with the  PTP trading signal for the Wheat continuous futures contract. Here, we use leverage to target a daily volatility of 1, for clarity. We have provided a 95\% confidence interval in the top plot to illustrate our predictive standard deviation.}
    \label{fig:X-Trend-g-positions}
\end{figure}

\begin{figure}[t]
    \centering
    \includegraphics[width=\textwidth]{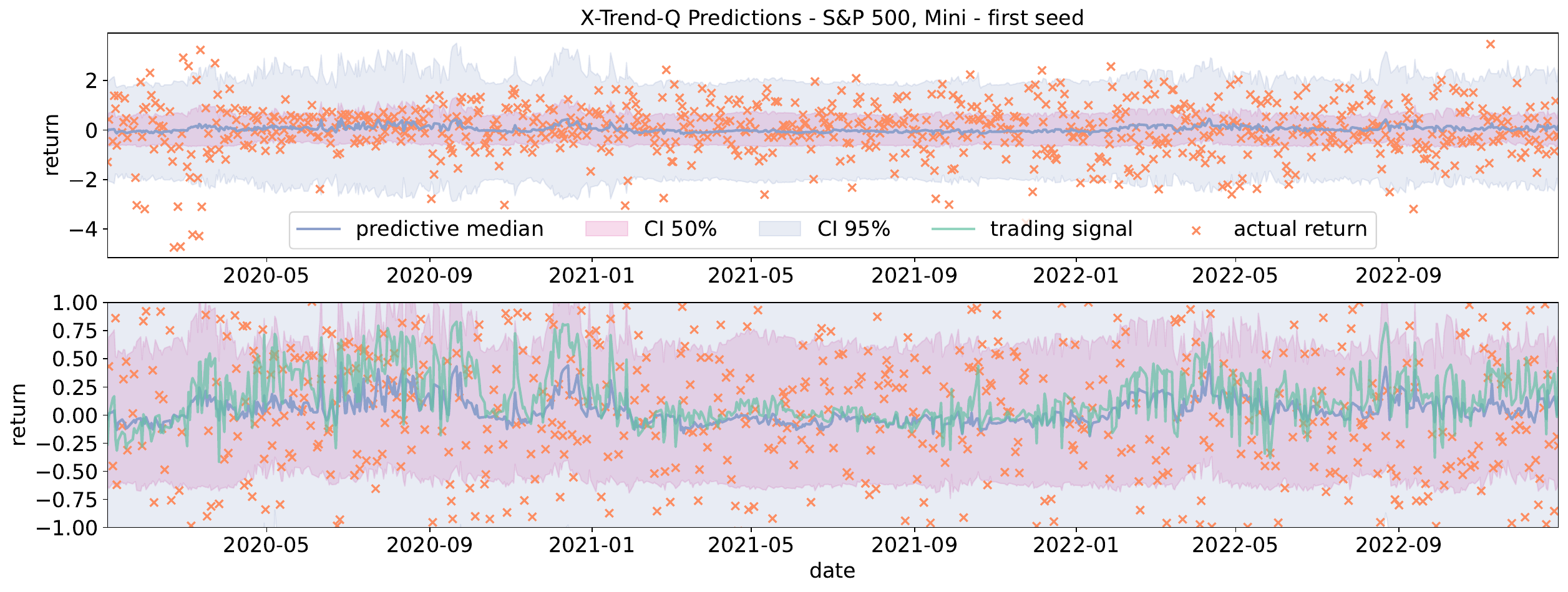}
    \caption{The relationship between X-Trend-Q predictive median, inter-quartile range, and 95\% confidence interval with the PTP trading signal for S\&P 500 Mini continuous futures contract. The bottom plot is a zoomed version of the top plot, with the trading signal superimposed. Here, we target daily volatility of 1, for clarity. 
    }
    \label{fig:X-Trend-q-positions}
\end{figure}

In~\cref{fig:X-Trend-g-positions} and ~\cref{fig:X-Trend-q-positions} we explore the relationship between our predictive returns output distribution and the $\mathrm{PTP}$ trading signal. In both examples, we observe that the predictive mean and median only make small movements above and below zero. This indicates that the model has learnt that we are operating in a low signal-to-noise environment, aiming to capitalize on slight market inefficiencies. Despite observing that there are small spikes and dips in predictive volatility, the estimate typically remains close to $1$, which suggests that our volatility targeting step, which targets $1$ in this example, is functioning as expected.

For X-Trend-G (\cref{fig:X-Trend-g-positions}), we observe that there is an almost linear relationship between the predictive mean and trading signal, indicating that the strategy uses the predictive mean as a measure of conviction and sizes the position accordingly. The relationship between predictive volatility and position is less clear. This suggests that it is the predictive mean driving the strategy and the predictive volatility is being used for volatility scaling. We can observe that the volatility scaling pre-processing job of altering leverage based on the $60$-day ex-ante volatility is doing a good job because we can observe that the $95$\% confidence interval is fairly stable around $\pm 1.96$ when targeting a daily volatility of $1$.

For X-Trend-Q (\cref{fig:X-Trend-q-positions}), we observe that the trading signal deviates more from the predictive median than X-Trend-G, indicating that it incorporates the full predictive distribution into the trading signal. This is likely the reason that the Sharpe for X-Trend-Q outperforms X-Trend-G by $7.1$\% in the few-shot setting over the turbulent period of $2018$ to $2023$. This supports our hypothesis that, for the best results, Gaussianity of returns should not be assumed in the few-shot setting. The assumption of Gaussian returns may be more appropriate in the low-resource setting.

\begin{figure}[t]
    \centering
    \includegraphics[width=\textwidth]{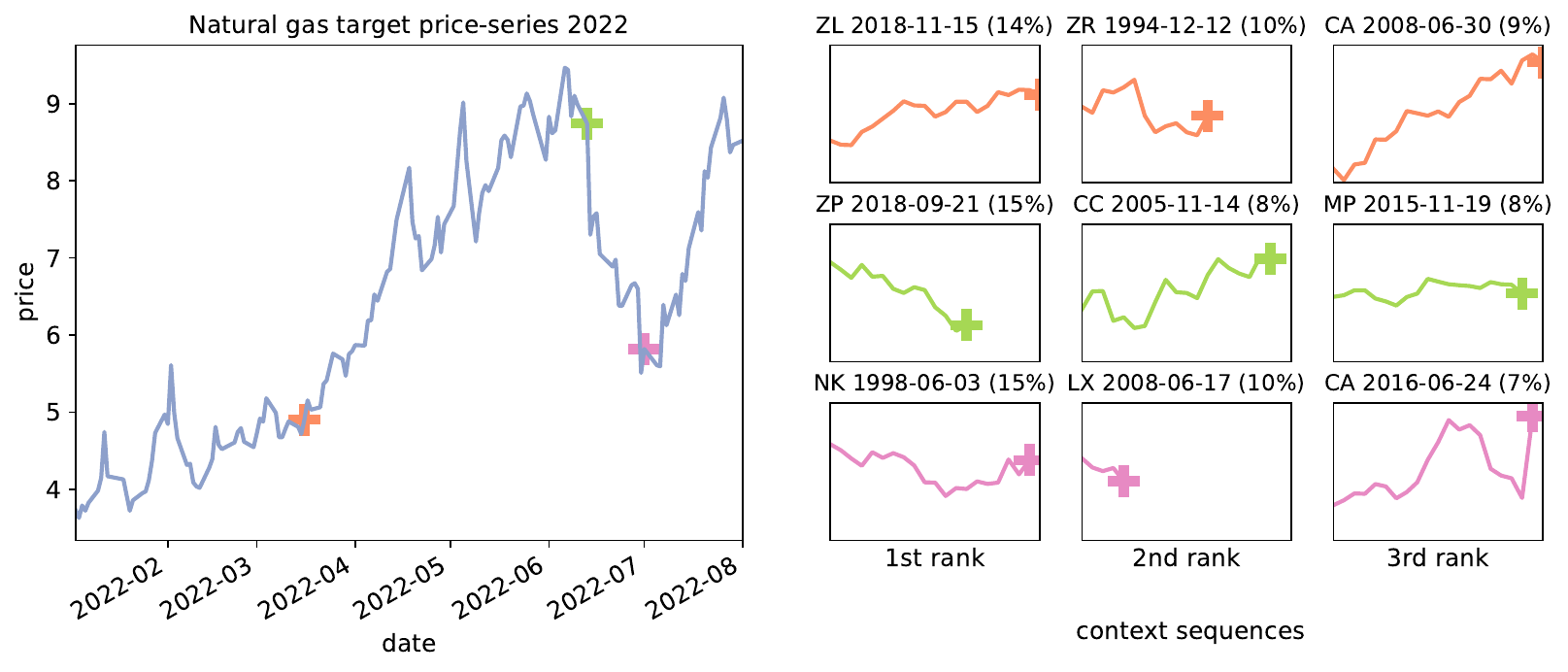}
    \caption{An illustrative example of the top $3$ attention weights for the target natural gas futures contract in 2022, a period exhibiting significant trends due to the Russia-Ukraine conflict. The $+$ symbols align the query (target) with associated hidden states for keys in the context sets, using colours to match attention weights with the time of forecast. The points we focus on are: the beginning of a significant uptrend (top), the beginning of a significant downtrend (middle), and a reversion (bottom). This example uses X-Trend with change-point segmented context sequences, $|\mathcal{C}|=20$ (meaning a uniform attention pattern would set all weights to $5\%$), and max length $l_{\mathrm{max}}=20$. We list the ticker, hidden-state date, and attention weight for each context sequence. The context contracts plotted are Soybean Oil (ZL), Rough Rice (ZR), CAC 40 index (CA), Platinum (ZP), Cocoa (CC), Mexican Peso (MP), Nikkei index (NK) and FTSE 100 index (LX).}
    \label{fig:att-weights}
\end{figure}

We provide an illustrative example of how we can interpret the cross-attention weights in~\cref{fig:att-weights} for the natural gas futures contract in $2022$; a period exhibiting significant trends due to the Russia-Ukraine conflict. We examine $3$ points in time, which correspond to $3$ different regimes, and in all cases, the top $3$ attention weights are highly intuitive. The target point at the beginning of an uptrend correctly identifies another commodity uptrend with the highest weight, with the other top $2$ being a commodity mean-reversion and a large equity uptrend. The target point at the beginning of the large downtrend clearly identifies another commodity sequence with a large downtrend, with almost double the weighting of the next highest. The target point during the beginning of a reversal identifies a slight downtrend with reversion, an extremely short downtrend, and an uptrend with significant reversion for the top three weights. Interestingly, in this case, the model actually identifies similarities with equities sequences instead of commodities.

\section{Related Works}
\textbf{Transfer Learning.} We can transfer features learned from one dataset to enable learning from new datasets more quickly. Transferring NN weights can also enable learning from new smaller datasets. This can be done by \emph{fine-tuning} the weights of a NN on a new task with SGD or \emph{linear-probing} which freezes lower layers of the network and only updates the top layers~\cite{FineTuningVsLinearProbing}.

\textbf{Few-shot learning.} Few-shot learning is characterized by enabling NNs to be able to learn to make predictions using little data. One key idea for enabling this is to train the few-shot learning agent in the same way that it is used for testing. This is the idea of \emph{episodic learning} where if we wish to train a model to make predictions using only $k$ examples or \emph{shots} then we need to train the model by performing $k$-shot learnings~\cite{vinyals2016matching, finn2017model}. 
Non-parametric methods are a natural fit for few-shot learning~\cite{patacchiola2020bayesian}. Learning distances between a few context points and a new target point has been fruitful for few-shot image classification~\cite{snell2017prototypical, sung2018learning, chen2019closer}. Neural processes learn to sample functions like Gaussian Process~\cite{garnelo2018conditional, garnelo2018neural, kim2019attentive}. Automatic context construction for few-shot learning using change-point detection methods has also been explored for image datasets~\cite{harrison2020continuous}. Recently there have been works that challenge some of the common assumptions for few-shot image classification~\cite{tian2020rethinking, dhillon2019baseline, laenen2021episodes}.

\textbf{Few-shot learning for time-series.} Neural Processes~\cite{garnelo2018neural} which parameterize a distribution over functions have been employed for few-shot learning for time-series forecasting~\cite{willi2019recurrent}. Neural processes algorithmically use a latent variable to enable drawing a new function for each point by sampling; the latent variable can employ autoregressive transition dynamics for time-series~\cite{singh2019sequential}. One can condition the target time-series predictions on a context of time-series with a cross-attention mechanism~\cite{qin2019recurrent}. The cross-attention mechanism has been shown to be very effective for Neural Processes~\cite{kim2019attentive}. Gradient-based few-shot learning has also been successfully employed for time-series~\cite{woo2022deeptime}. From the literature and from our own initial experiments on toy data~(\Cref{sec:gp_draws_exp}) we found the cross-attention very effective for few-shot learning for time-series.

\textbf{Deep-learning for financial time-series}.
\label{sec:deep-learning-ts-related}
Building on the work of the vanilla DMNs, \cite{lim2019enhancing}, the \emph{Momentum Transformer}~\cite{wood2021trading}, which is a variant of the Temporal Fusion Transformer~\cite{TFT}, incorporates an attention mechanism to attend to prior LSTM hidden-states from the same sequence. The attention pattern naturally segments the time-series into regimes, where significant importance is placed on the final hidden state of each regime, motivating the cross-attention step in this paper. Other works demonstrate that causal convolutional filters can be be utilized to automatically generate features~\cite{jiang2020re, DeepInception}, as an alternative to the momentum factor approach used in this paper.  The work of \cite{zhang2019deeplob} utilizes convolutions followed by an LSTM to generate features from limit order book (LOB) data. There is growing evidence to suggest that we can benefit from a cross-section of assets to assist our forecasting, where the work by \cite{DeepInception} implements convolutional filters across assets and \cite{pu2023network} uses a graph learning model to reveal momentum spillover. Meta-learning has been used in finance to construct a partial index portfolio to track a benchmark index where the asset allocation is meta-learned~\cite{yang2023partial}.

For a broad review of the financial machine learning literature, we direct readers to~\cite{kelly2023financial}. For a more general overview of deep-learning techniques for time-series forecasting we direct readers to \cite{lim2021time}. 

\section{Conclusions and Future Work}
We introduce \textbf{X-Trend}: the \textbf{Cross} Attentive Time-Series \textbf{Trend} Network. It leverages few-shot learning and change-point detection to enable adaptation to new financial regimes. We show that it is able to recover from the COVID-19 draw-down almost twice as quickly as an equivalent neural time-series agent. Over the $5$-year period $2018$ to $2023$, we are able to improve risk-adjusted returns by $18.9\%$ compared to the baseline agent and around 10-fold compared to a conventional Time-series Momentum (TSMOM) strategy. This boost in performance is largely driven by our cross-attention step which transfers trends, from similar patterns in a context set. Furthermore, our model can generate profitable zero-shot trading signals in an extremely challenging low-resource setting where we trade a previously unseen asset we achieve a Sharpe of $0.47$ compared to loss-making time-series momentum baselines, both deep-learning based and conventional TSMOM. In this work we withheld assets from a standard dataset to test zero-shot performance; a future avenue of work could be applying this framework to an emerging asset class such as cryptocurrencies.

We illustrate the importance of constructing a good context set, where we improve Sharpe by $11.3\%$ after segmenting the sequences with change-point detection. A future avenue of work would be to further investigate context set construction. One possibility could be considering a cross-sectional approach where we attend across a universe of assets at the same time, motivated by the works~\cite{DeepInception, tan2023spatio} with the option of including lead-lag~\cite{zhang2023robust}. We could consider generating synthetic data for the context set \cite{wiese2020quant}. Another direction of work, inspired by the Neural Process literature, is to reconcile optimizing a Sharpe ratio with optimizing the evidence lower bound for variational inference required for latent variable Neural Process time-series models~\cite{qin2019recurrent, willi2019recurrent}. Finally, this work could be combined with other innovations that expand upon the standard Deep Momentum Network framework such as bringing transaction costs into the loss function~\cite{lim2019enhancing} and including change-point features in the decoder~\cite{wood2022slow}. We could also use self-attention in the temporal dimension \cite{wood2021trading, TFT}, introducing it in the decoder, and automatically generating features from our assets~\cite{DeepInception}.

\section{Acknowledgements}
The authors would like to thank the Oxford-Man Institute of Quantitative Finance for its generous support.
SR would like to thank the U.K. Royal Academy of Engineering.

\bibliography{refs}

\begin{thebibliography}{10}

\bibitem{MomentumCrashes}
Kent Daniel and Tobias~J. Moskowitz.
\newblock Momentum crashes.
\newblock {\em Journal of Financial Economics}, 122(2):221 -- 247, 2016.

\bibitem{MomentumTurningPoints}
Ashish Garg, Christian~L Goulding, Campbell~R Harvey, and Michele Mazzoleni.
\newblock Momentum turning points.
\newblock {\em Available at SSRN 3489539}, 2021.

\bibitem{lim2019enhancing}
Bryan Lim, Stefan Zohren, and Stephen Roberts.
\newblock Enhancing time-series momentum strategies using deep neural networks.
\newblock {\em The Journal of Financial Data Science}, 1(4):19--38, 2019.

\bibitem{wood2022slow}
Kieran Wood, Stephen Roberts, and Stefan Zohren.
\newblock Slow momentum with fast reversion: A trading strategy using deep learning and changepoint detection.
\newblock {\em The Journal of Financial Data Science}, 4(1):111--129, 2022.

\bibitem{moskowitz2012time}
Tobias~J Moskowitz, Yao~Hua Ooi, and Lasse~Heje Pedersen.
\newblock Time series momentum.
\newblock {\em Journal of financial economics}, 104(2):228--250, 2012.

\bibitem{FactorCrowding}
Nick Baltas.
\newblock The impact of crowding in alternative risk premia investing.
\newblock {\em Financial Analysts Journal}, 75(3):89--104, 2019.

\bibitem{brown2022crowded}
Gregory~W Brown, Philip Howard, and Christian~T Lundblad.
\newblock Crowded trades and tail risk.
\newblock {\em The Review of Financial Studies}, 35(7):3231--3271, 2022.

\bibitem{vinyals2016matching}
Oriol Vinyals, Charles Blundell, Timothy Lillicrap, Daan Wierstra, et~al.
\newblock Matching networks for one shot learning.
\newblock {\em Advances in neural information processing systems}, 29, 2016.

\bibitem{ravi2016optimization}
Sachin Ravi and Hugo Larochelle.
\newblock Optimization as a model for few-shot learning.
\newblock 2016.

\bibitem{snell2017prototypical}
Jake Snell, Kevin Swersky, and Richard Zemel.
\newblock Prototypical networks for few-shot learning.
\newblock {\em Advances in neural information processing systems}, 30, 2017.

\bibitem{duan2016rl}
Yan Duan, John Schulman, Xi~Chen, Peter~L Bartlett, Ilya Sutskever, and Pieter Abbeel.
\newblock Rl$^2$: Fast reinforcement learning via slow reinforcement learning.
\newblock {\em arXiv preprint arXiv:1611.02779}, 2016.

\bibitem{wang2016learning}
Jane~X Wang, Zeb Kurth-Nelson, Dhruva Tirumala, Hubert Soyer, Joel~Z Leibo, Remi Munos, Charles Blundell, Dharshan Kumaran, and Matt Botvinick.
\newblock Learning to reinforcement learn.
\newblock {\em arXiv preprint arXiv:1611.05763}, 2016.

\bibitem{finn2017model}
Chelsea Finn, Pieter Abbeel, and Sergey Levine.
\newblock Model-agnostic meta-learning for fast adaptation of deep networks.
\newblock In {\em International Conference on Machine Learning}, pages 1126--1135. PMLR, 2017.

\bibitem{poterba1988mean}
James~M Poterba and Lawrence~H Summers.
\newblock Mean reversion in stock prices: Evidence and implications.
\newblock {\em Journal of financial economics}, 22(1):27--59, 1988.

\bibitem{vayanos2013institutional}
Dimitri Vayanos and Paul Woolley.
\newblock An institutional theory of momentum and reversal.
\newblock {\em The Review of Financial Studies}, 26(5):1087--1145, 2013.

\bibitem{SharpeRatio}
William~F. Sharpe.
\newblock The sharpe ratio.
\newblock {\em The Journal of Portfolio Management}, 21(1):49--58, 1994.

\bibitem{RamaContUniversality}
Justin Sirignano and Rama Cont.
\newblock Universal features of price formation in financial markets: Perspectives from deep learning.
\newblock {\em SSRN}, 2018.

\bibitem{zhang2019deeplob}
Zihao Zhang, Stefan Zohren, and Stephen Roberts.
\newblock Deeplob: Deep convolutional neural networks for limit order books.
\newblock {\em IEEE Transactions on Signal Processing}, 67(11):3001--3012, 2019.

\bibitem{wood2021trading}
Kieran Wood, Sven Giegerich, Stephen Roberts, and Stefan Zohren.
\newblock Trading with the momentum transformer: An intelligent and interpretable architecture.
\newblock {\em arXiv preprint arXiv:2112.08534}, 2021.

\bibitem{kim2019attentive}
Hyunjik Kim, Andriy Mnih, Jonathan Schwarz, Marta Garnelo, Ali Eslami, Dan Rosenbaum, Oriol Vinyals, and Yee~Whye Teh.
\newblock Attentive neural processes.
\newblock {\em arXiv preprint arXiv:1901.05761}, 2019.

\bibitem{doersch2020crosstransformers}
Carl Doersch, Ankush Gupta, and Andrew Zisserman.
\newblock Crosstransformers: spatially-aware few-shot transfer.
\newblock {\em Advances in Neural Information Processing Systems}, 33:21981--21993, 2020.

\bibitem{vaswani2017attention}
Ashish Vaswani, Noam Shazeer, Niki Parmar, Jakob Uszkoreit, Llion Jones, Aidan~N Gomez, {\L}ukasz Kaiser, and Illia Polosukhin.
\newblock Attention is all you need.
\newblock {\em Advances in neural information processing systems}, 30, 2017.

\bibitem{garnett2010sequential}
Roman Garnett, Michael~A Osborne, Steven Reece, Alex Rogers, and Stephen~J Roberts.
\newblock Sequential bayesian prediction in the presence of changepoints and faults.
\newblock {\em The Computer Journal}, 53(9):1430--1446, 2010.

\bibitem{saatcci2010gaussian}
Yunus Saat{\c{c}}i, Ryan~D Turner, and Carl~E Rasmussen.
\newblock Gaussian process change point models.
\newblock In {\em Proceedings of the 27th International Conference on Machine Learning (ICML-10)}, pages 927--934, 2010.

\bibitem{TFT}
B~Lim, SO~Arik, N~Loeff, and T~Pfister.
\newblock Temporal fusion transformers for interpretable multi-horizon time series forecasting. arxiv.
\newblock {\em arXiv preprint arXiv:1912.09363}, 2019.

\bibitem{TSMomAndVolScaling}
Abby~Y. Kim, Yiuman Tse, and John~K. Wald.
\newblock Time series momentum and volatility scaling.
\newblock {\em Journal of Financial Markets}, 30:103 -- 124, 2016.

\bibitem{VolTargeting}
Campbell~R. Harvey, Edward Hoyle, Russell Korgaonkar, Sandy Rattray, Matthew Sargaison, and Otto van Hemert.
\newblock The impact of volatility targeting.
\newblock {\em SSRN}, 2018.

\bibitem{robbins1951stochastic}
Herbert Robbins and Sutton Monro.
\newblock A stochastic approximation method.
\newblock {\em The annals of mathematical statistics}, pages 400--407, 1951.

\bibitem{DeepLearningBook}
Ian Goodfellow, Yoshua Bengio, and Aaron Courville.
\newblock {\em Deep Learning}.
\newblock MIT Press, 2016.
\newblock \url{http://www.deeplearningbook.org}.

\bibitem{CenturyOfTrendFollowing}
Brian Hurst, Yao~Hua Ooi, and Lasse~Heje Pedersen.
\newblock A century of evidence on trend-following investing.
\newblock {\em The Journal of Portfolio Management}, 44(1):15--29, 2017.

\bibitem{AHLMomentum}
Jamil Baz, Nicolas Granger, Campbell~R. Harvey, Nicolas Le~Roux, and Sandy Rattray.
\newblock Dissecting investment strategies in the cross section and time series.
\newblock {\em SSRN}, 2015.

\bibitem{DeepInception}
Tom Liu, Stephen Roberts, and Stefan Zohren.
\newblock Deep inception networks: A general end-to-end framework for multi-asset quantitative strategies.
\newblock {\em arXiv preprint arXiv:2307.05522}, 2023.

\bibitem{hochreiter1997long}
Sepp Hochreiter and J{\"u}rgen Schmidhuber.
\newblock Long short-term memory.
\newblock {\em Neural computation}, 9(8):1735--1780, 1997.

\bibitem{EntityEmbeddings}
Cheng Guo and Felix Berkhahn.
\newblock Entity embeddings of categorical variables.
\newblock {\em arXiv preprint arXiv:1604.06737}, 2016.

\bibitem{clevert2015fast}
Djork-Arn{\'e} Clevert, Thomas Unterthiner, and Sepp Hochreiter.
\newblock Fast and accurate deep network learning by exponential linear units (elus).
\newblock {\em arXiv preprint arXiv:1511.07289}, 2015.

\bibitem{wen2017multi}
Ruofeng Wen, Kari Torkkola, Balakrishnan Narayanaswamy, and Dhruv Madeka.
\newblock A multi-horizon quantile recurrent forecaster.
\newblock {\em arXiv preprint arXiv:1711.11053}, 2017.

\bibitem{garnelo2018neural}
Marta Garnelo, Jonathan Schwarz, Dan Rosenbaum, Fabio Viola, Danilo~J Rezende, SM~Eslami, and Yee~Whye Teh.
\newblock Neural processes.
\newblock {\em arXiv preprint arXiv:1807.01622}, 2018.

\bibitem{qin2019recurrent}
Shenghao Qin, Jiacheng Zhu, Jimmy Qin, Wenshuo Wang, and Ding Zhao.
\newblock Recurrent attentive neural process for sequential data.
\newblock {\em arXiv preprint arXiv:1910.09323}, 2019.

\bibitem{FineTuningVsLinearProbing}
Ananya Kumar, Aditi Raghunathan, Robbie Jones, Tengyu Ma, and Percy Liang.
\newblock Fine-tuning can distort pretrained features and underperform out-of-distribution.
\newblock {\em arXiv preprint arXiv:2202.10054}, 2022.

\bibitem{patacchiola2020bayesian}
Massimiliano Patacchiola, Jack Turner, Elliot~J Crowley, Michael O'Boyle, and Amos~J Storkey.
\newblock Bayesian meta-learning for the few-shot setting via deep kernels.
\newblock {\em Advances in Neural Information Processing Systems}, 33:16108--16118, 2020.

\bibitem{sung2018learning}
Flood Sung, Yongxin Yang, Li~Zhang, Tao Xiang, Philip~HS Torr, and Timothy~M Hospedales.
\newblock Learning to compare: Relation network for few-shot learning.
\newblock In {\em Proceedings of the IEEE conference on computer vision and pattern recognition}, pages 1199--1208, 2018.

\bibitem{chen2019closer}
Wei-Yu Chen, Yen-Cheng Liu, Zsolt Kira, Yu-Chiang~Frank Wang, and Jia-Bin Huang.
\newblock A closer look at few-shot classification.
\newblock {\em arXiv preprint arXiv:1904.04232}, 2019.

\bibitem{garnelo2018conditional}
Marta Garnelo, Dan Rosenbaum, Christopher Maddison, Tiago Ramalho, David Saxton, Murray Shanahan, Yee~Whye Teh, Danilo Rezende, and SM~Ali Eslami.
\newblock Conditional neural processes.
\newblock In {\em International Conference on Machine Learning}, pages 1704--1713. PMLR, 2018.

\bibitem{harrison2020continuous}
James Harrison, Apoorva Sharma, Chelsea Finn, and Marco Pavone.
\newblock Continuous meta-learning without tasks.
\newblock {\em Advances in neural information processing systems}, 33:17571--17581, 2020.

\bibitem{tian2020rethinking}
Yonglong Tian, Yue Wang, Dilip Krishnan, Joshua~B Tenenbaum, and Phillip Isola.
\newblock Rethinking few-shot image classification: a good embedding is all you need?
\newblock In {\em Computer Vision--ECCV 2020: 16th European Conference, Glasgow, UK, August 23--28, 2020, Proceedings, Part XIV 16}, pages 266--282. Springer, 2020.

\bibitem{dhillon2019baseline}
Guneet~S Dhillon, Pratik Chaudhari, Avinash Ravichandran, and Stefano Soatto.
\newblock A baseline for few-shot image classification.
\newblock {\em arXiv preprint arXiv:1909.02729}, 2019.

\bibitem{laenen2021episodes}
Steinar Laenen and Luca Bertinetto.
\newblock On episodes, prototypical networks, and few-shot learning.
\newblock {\em Advances in Neural Information Processing Systems}, 34:24581--24592, 2021.

\bibitem{willi2019recurrent}
Timon Willi, Jonathan Masci, J{\"u}rgen Schmidhuber, and Christian Osendorfer.
\newblock Recurrent neural processes.
\newblock {\em arXiv preprint arXiv:1906.05915}, 2019.

\bibitem{singh2019sequential}
Gautam Singh, Jaesik Yoon, Youngsung Son, and Sungjin Ahn.
\newblock Sequential neural processes.
\newblock {\em Advances in Neural Information Processing Systems}, 32, 2019.

\bibitem{woo2022deeptime}
Gerald Woo, Chenghao Liu, Doyen Sahoo, Akshat Kumar, and Steven Hoi.
\newblock Deeptime: Deep time-index meta-learning for non-stationary time-series forecasting.
\newblock {\em arXiv preprint arXiv:2207.06046}, 2022.

\bibitem{jiang2020re}
Jingwen Jiang, Bryan~T Kelly, and Dacheng Xiu.
\newblock (re-) imag (in) ing price trends.
\newblock {\em Chicago Booth Research Paper}, (21-01), 2020.

\bibitem{pu2023network}
Xingyue~Stacy Pu, Stephen Roberts, Xiaowen Dong, and Stefan Zohren.
\newblock Network momentum across asset classes.
\newblock {\em Stephen and Dong, Xiaowen and Zohren, Stefan, Network Momentum across Asset Classes (August 7, 2023)}, 2023.

\bibitem{yang2023partial}
Yongxin Yang and Timothy Hospedales.
\newblock Partial index tracking: A meta-learning approach.
\newblock In {\em Conference on Lifelong Learning Agents}, pages 415--436. PMLR, 2023.

\bibitem{kelly2023financial}
Bryan~T Kelly and Dacheng Xiu.
\newblock Financial machine learning.
\newblock Technical report, National Bureau of Economic Research, 2023.

\bibitem{lim2021time}
Bryan Lim and Stefan Zohren.
\newblock Time-series forecasting with deep learning: a survey.
\newblock {\em Philosophical Transactions of the Royal Society A}, 379(2194):20200209, 2021.

\bibitem{tan2023spatio}
Wee~Ling Tan, Stephen Roberts, and Stefan Zohren.
\newblock Spatio-temporal momentum: Jointly learning time-series and cross-sectional strategies.
\newblock {\em arXiv preprint arXiv:2302.10175}, 2023.

\bibitem{zhang2023robust}
Yichi Zhang, Mihai Cucuringu, Alexander~Y Shestopaloff, and Stefan Zohren.
\newblock Robust detection of lead-lag relationships in lagged multi-factor models.
\newblock {\em arXiv preprint arXiv:2305.06704}, 2023.

\bibitem{wiese2020quant}
Magnus Wiese, Robert Knobloch, Ralf Korn, and Peter Kretschmer.
\newblock Quant gans: deep generation of financial time series.
\newblock {\em Quantitative Finance}, 20(9):1419--1440, 2020.

\bibitem{kim2018bayesian}
Taesup Kim, Jaesik Yoon, Ousmane Dia, Sungwoong Kim, Yoshua Bengio, and Sungjin Ahn.
\newblock {Bayesian model-agnostic meta-learning}.
\newblock {\em arXiv preprint arXiv:1806.03836}, 2018.

\bibitem{kingma2013auto}
Diederik~P Kingma and Max Welling.
\newblock Auto-encoding variational bayes.
\newblock {\em arXiv preprint arXiv:1312.6114}, 2013.

\bibitem{Williams89alearning}
Ronald~J. Williams and David Zipser.
\newblock A learning algorithm for continually running fully recurrent neural networks, 1989.

\bibitem{ADAM}
Diederik Kingma and Jimmy Ba.
\newblock Adam: A method for stochastic optimization.
\newblock In {\em International Conference on Learning Representations {(ICLR)}}, 2015.

\bibitem{dropout}
Nitish Srivastava, Geoffrey Hinton, Alex Krizhevsky, Ilya Sutskever, and Ruslan Salakhutdinov.
\newblock Dropout: A simple way to prevent neural networks from overfitting.
\newblock {\em Journal of Machine Learning Research}, 15:1929--1958, 2014.

\bibitem{pytorch}
Adam Paszke, Sam Gross, Soumith Chintala, Gregory Chanan, Edward Yang, Zachary DeVito, Zeming Lin, Alban Desmaison, Luca Antiga, and Adam Lerer.
\newblock Automatic differentiation in {PyTorch}.
\newblock In {\em Autodiff Workshop -- Advances in Neural Information Processing {(NeurIPS)}}, 2017.

\end{thebibliography}
\bibliographystyle{unsrt}

\clearpage

\begin{appendices}

\crefalias{section}{appsec}
\crefalias{subsection}{appsec}
\crefalias{subsubsection}{appsec}

\onecolumn

\section*{\LARGE Supplementary Material}
\label{sec:appendix}
\section{Gaussian Process Change Point Segmentation}
\label{app:cpd}
Even after using returns to linearly de-trending the price-series and by employing volatility scaling to target a consistent volatility, we still encounter occasional and significant periods of disequilibrium or regime change. The work in~\cite{wood2022slow} explores how a Gaussian Process (GP) based online change-point detection (CPD) ~\cite{garnett2010sequential, saatcci2010gaussian} module can be inserted to help address any transitions in regime, which motivates the episodic approach taken in this paper. The work by ~\cite{wood2022slow}, assumes that there is a single change-point in some pre-specified lookback window (LBW), then calculates the change-point location and severity, therefore an approximately stationary regime will correspond with a very low severity. The severity is computed by the improvement in log marginal-likelihood using a change-point covariance kernel, in comparison to a simple Gaussian Processes, $\mathcal{GP}(\cdot, \cdot)$\footnote{A typical choice could be an Ornstein–Uhlenbeck process, which is the Matérn Kernel 1/2 kernel. For consistency with the motivating work~\cite{wood2022slow}, we use the Matérn 3/2 kernel.}. The Change-point kernel assumes that there are instead two underlying Gaussian Processes of the same kernel, $\mathcal{GP}_1(\cdot, \cdot)$ and $\mathcal{GP}_2(\cdot, \cdot)$ (which we refer to together as $\mathcal{GP}_{\mathrm{C}}$), either side of the assumed change-point with a soft transition from the first to the second. That is, we measure the benefit of using two separate kernels instead of one. We leave the location of the transition $t_{\mathrm{CPD}}$ as a free variable, which we tune when we maximize the likelihood. In this work we use a LBW of $l_{\mathrm{lbw}}=21$ as a compromise between speed of detection and robustness to noise~\cite{wood2022slow}. This is not to be confused with the fact that we can have regimes longer than $l_{\mathrm{lbw}}$ because we identify regime change when the severity threshold $\nu$ is reached. We detail our CPD algorithm in~\cref{alg:cpd}. We use a change-point severity of $\nu=0.9$ for the experiments where we set maximum regime length as $l_{\mathrm{max}}=21$ and we set $\nu=0.95$ for the experiments where we use $l_{\mathrm{max}}=63$. We disregard regimes with lengths less than $l_\min=5$.

\begin{algorithm}[H]
    \SetAlgoLined
    \KwData{price series $p_{1:T}^{(i)}$, CPD LBW $l_{\mathrm{lbw}}$, CPD threshold $\nu$, min. segment length $l_\min$ , max. segment length $l_\max$}
    \KwResult{segmented price series $\{p^{(i)}_{t_0:t_1}\}_{(t_0, t_1)\in \mathcal{R}}$}
    Initialize: $t \gets T$, $t_1 \gets T$, regimes $\mathcal{R} \gets \emptyset$ \;
    \While{$t \geq 0$}{
        Fit $\mathcal{GP}$ with Matérn 3/2 kernel on $p_{-l_{\mathrm{lbw}}:t}$ and calculate marginal likelihood, $L_{M}$\;
        Fit $\mathcal{GP}_{\mathrm{C}}$ with Change-point kernel on $p_{-l_{\mathrm{lbw}}:t}$ and calculate marginal likelihood, $L_{C}$ and change-point location hyperparameter $t_{\mathrm{CPD}}$\;
        \eIf{$\frac{L_C}{L_M + L_C} \geq \nu$}{
            $t_0 \gets \lceil t_c \rceil $ \;
            \If{$t_1 - t_0 \geq l_\min$}{ $\mathcal{R} \gets \mathcal{R} \cup \{(t_0, t_1)\}\;$}
            $t \gets \lfloor t_c \rfloor - 1 $ \# so that `good' representation isn't corrupted\;
             $t_1 \gets t$
        }{
            $t\gets t-1$\;
            \If{$t_1-t>l_\max$}{
            $t \gets t_1-l_\max$\;}
            \If{$t_1-t=l_\max$}{
            $\mathcal{R} \gets \mathcal{R} \cup \{(t, t_1)\}$\;
            $t_1 \gets t$\;}
        }
    }
    \caption{Time-series CPD segmentation \label{alg:cpd}}
\end{algorithm}

\begin{figure}
    \centering
    \includegraphics[width=1.0\textwidth]{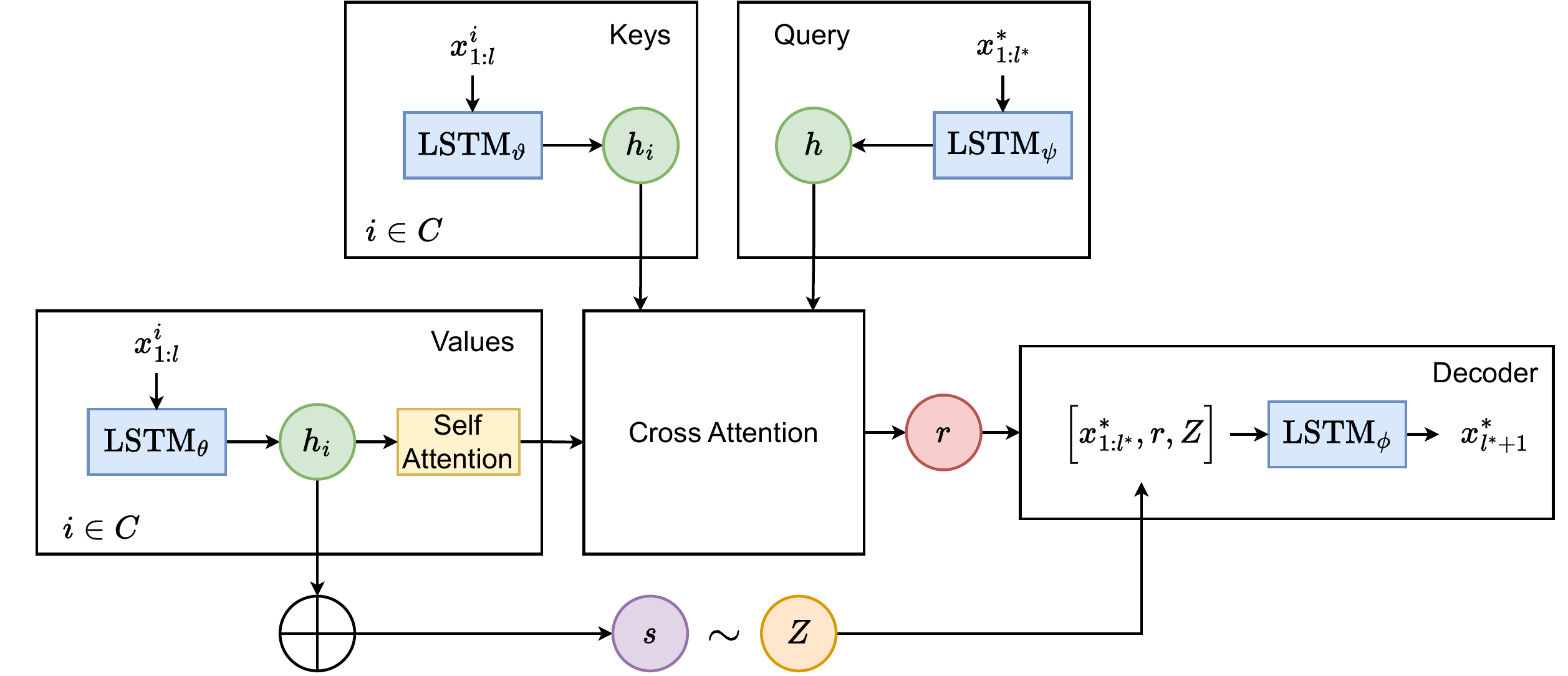}
    \caption{Recurrent attentive neural process. The keys and values are the context set hidden states. We have a self-attention mechanism over hidden states in the values. The LSTM encoders for the values and keys have separate parameters. The queries are encoded by a separate LSTM with parameters $\psi$. The cross-attention mechanism outputs a similarity between the target $x^*$ and the context $\mathcal{C}$ representation of the context set: $r$ and the latent variable $s$ is sampled to produce a summary of the context set to condition an LSTM decoder to make predictions on a target set $x^*$.}
    \label{fig:rec_np}
\end{figure}

\section{Recurrent Attentive Neural Process Experiments}
\label{sec:gp_draws_exp}
\subsection{Model Architecture}
We use the recurrent attentive neural process~\cite{qin2019recurrent} as a benchmark to study the importance of certain model components. In this section, we will summarize the model and provide an overview in~\cref{fig:rec_np}.

We use $|\mathcal{C}|$ contexts which are sequences of length $l$. The targets $x^*_{1:k}$ have a length $k$ and are required to produce a prediction $y*:=x^*_{k+1}$. Causally the contexts are all observed at time-steps prior to the targets. The contexts are passed into an LSTM encoder with parameters $\theta$. The final hidden states $\mathbf{h}_i$ are passed into a self-attention module over contexts before the cross-attention branch.

We perform cross attention between the context encodings $\mathbf{h}_i$ and the targets $x^*_{1:k}$. The targets are encoded by a separate LSTM network with parameters $\psi$ and the final hidden state $h$ is the query vector in the cross-attention model. The encoded contexts $\mathbf{h}_i, \forall i \in C$ are the values. The keys are the contexts, which are encoded similarly to the values but with another LSTM encoder with parameters $\varphi$. The output of the cross attention is $r$ and is used to condition the decoder~\cite{kim2018bayesian, qin2019recurrent}.

The encodings $\mathbf{h}_i$ from the encoder are then aggregated with a sum operation $\bigoplus$ in~\cref{fig:rec_np}. These encodings are then passed into two separate linear layers which are used to parameterize the mean and variance of a latent variable $s$ which is distributed according to a Gaussian distribution and trained using the reparameterization trick~\cite{kingma2013auto}. The context encodings $r$ and the draw from $s$, $Z$ are  stacked on top of the targets $x^*$ before producing predictions with an LSTM decoder network with parameters $\phi$~\cite{qin2019recurrent, willi2019recurrent}. 

\subsection{Results}
We generate a dataset of GP draws for the recurrent attentive neural process~\cite{qin2019recurrent}~\cref{fig:rec_np} to produce forecasts. The GP curves are generated using an RBF (Radial Basis Function) kernel with a length scale of $0.4$ and noise variance of $1.0$. Each context is drawn from a different GP draw. The length of each sequence in the context $\mathcal{C}$, the input $x_c$ and output $y_c$ are temporally connected sequences of length $l_C \sim \textrm{U}[10, 30]$. Likewise, the target is taken from a separate GP draw where the input $x_T$ and the output $y_T$ are temporally connected sequences of length $l_T \sim \textrm{U}[10, 30]$. The target decoder is trained for $50$k iterations using teacher forcing and during testing unrolls the forecast using the previous prediction as a new input~\cite{Williams89alearning}. We use a Gaussian likelihood and assess test performance using the MSE.

\begin{figure}
    \centering
    \includegraphics[width=0.8\textwidth]{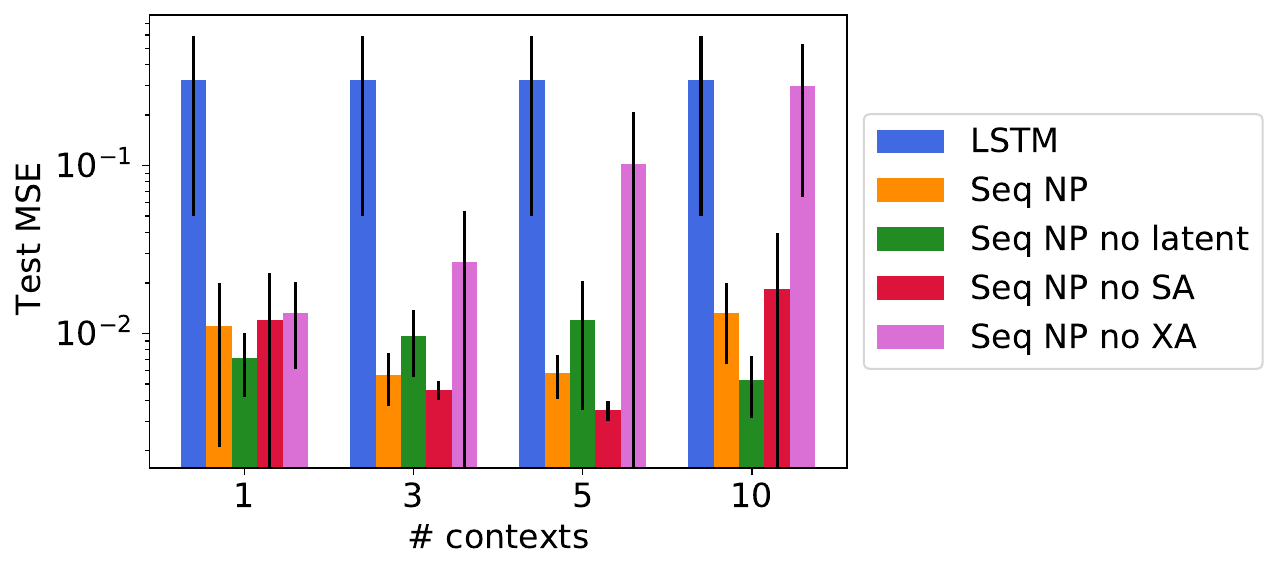}
    \caption{The final test MSE for different model choices for increasing number of context sequences. The results are a mean $\pm 1$ standard error over $5$ seeds.}
    \label{fig:final_test_mse_training_curves}
\end{figure}

From~\cref{fig:final_test_mse_training_curves} we can see that the baseline LSTM forecaster with no context $\mathcal{C}$, suffers from overfitting after training for a fixed number of iterations. The full sequential NP with latent variable, self-attention over the context hidden states, and cross-attention between target and contexts obtains good forecasting performance for all different experimental settings for different context set sizes. 

When we ablate away certain components, we see that performance remains the same when removing the latent variable and the self-attention. However, If we remove the cross-attention between the contexts and the target sequences then the sequential NP underfits severely. This underfitting becomes more severe as more context sequences are used to condition the sequential NP (\cref{fig:final_test_mse_training_curves}).

\section{Training Details}
\label{app:expt-details}
We calibrate our model using the training data by optimizing the Sharpe loss function via minibatch Stochastic Gradient Descent (SGD)~\cite{robbins1951stochastic}, using the \emph{Adam} optimizer \cite{ADAM}. We employ dropout~\cite{dropout}, which helps to prevent the model from overfitting by randomly removing hidden nodes in the training phase. We list the fixed model parameters for each architecture in~\cref{tab:fixed-params}, including early stopping patience. We keep the last 10\% of the training data, for each asset, as a validation set. We implement 10 iterations of random grid search, as an outer optimization loop, to select the best hyperparameters, based on the validation set. The hyperparameter search grid for each architecture is listed in~\cref{tab:hyperparams}. We perform 10 full repeats of the outer optimization loop and ensemble the 10 models for our experiments to reduce noise. 

Our model was implemented in the deep-learning framework \emph{PyTorch} \cite{pytorch}. It was trained on a \emph{NVIDIA GeForce RTX 3090} GPU.
\begin{table}
\begin{minipage}{.5\linewidth}
\centering
\centering
\caption{Hyperparameter Search Range.\label{tab:hyperparams}}
\begin{tabular}{@{}lll@{}}
\toprule
\textbf{Hyperparameters}           & \textbf{Random Grid}                                \\ \midrule
Dropout Rate                       & 0.3, 0.4, 0.5                                 
\\
Hidden Layer Size, $d_h$                 & 64, 128                                     
\\
Minibatch Size, $b$                     & 64, 128 \\

Max Gradient Norm                  & $10^{-2},~ 10^{0},~ 10^{2}$                        \\
\bottomrule
\end{tabular}

\end{minipage}
\begin{minipage}{.5\linewidth}
\centering
\caption{Training Fixed Parameters.\label{tab:fixed-params}}
\begin{tabular}{@{}lll@{}}
\toprule
\textbf{Parameter}           & \textbf{Value} \\ \midrule
Learning Rate                      & $10^{-3}$      \\
Target, training warm-up steps, $l_s$ & $63$ \\
Target, total LSTM steps, $l_t$ & $126$ \\
Early stopping patience & 10 \\
Maximum SGD iterations &  100 \\
Number attention heads &  4 \\
Attention dimension, $d_{\mathrm{att}}$ & $d_h$ \\
Joint loss weight, $\alpha_{\mathrm{X-Trend-G}}$ & 1\\
Joint loss weight, $\alpha_{\mathrm{X-Trend-Q}}$ & 5\\
\bottomrule
\end{tabular}
\end{minipage}
\end{table}

\section{Datasets}
\label{app:data}
We backtest our X-Trend variants on a portfolio of 50 liquid, continuous futures contracts over the period 1990--2023, extracted from the Pinnacle Data Corp CLC Database. The futures contracts are chained together using the backwards ratio-adjusted method. For our few-shot experiments, we use the same assets for both training and testing, using all assets in both~\cref{tab:dataset} and~\cref{tab:dataset-zeroshot}. For our zero-shot experiments, we randomly selected 20 of the 50 Pinnacle assets as the target set $\mathcal{I}_{ts}$, which we detail in ~\cref{tab:dataset-zeroshot}, leaving the other 30 for $\mathcal{I}_{tr}$, which we detail in~\cref{tab:dataset}. It should be noted that we chose to not include any fixed income contracts in the zero-shot portfolio, due to the fact that they are more correlated than the other futures contracts. 

\begin{table}[!htb]
\small
\begin{minipage}{.5\linewidth}
\centering
\caption{Assets in $\mathcal{I}_{tr}$ for zero-shot experiments and part of $\mathcal{I}$ for few-shot experiments. \label{tab:dataset}}
\begin{tabular}{ll}
\midrule
\textbf{Identifier} & \textbf{Description}        \\ \midrule
\multicolumn{2}{l}{{\underline{\textbf{Commodities (CM)}}}}\\
CC                  & COCOA                      \\ 
DA                  & MILK III, composite            \\


LB                  & LUMBER                     \\

SB                  & SUGAR \#11                 \\
ZA                  & PALLADIUM, electronic       \\
ZC                  & CORN, electronic           \\
ZF                  & FEEDER CATTLE, electronic  \\

ZI                  & SILVER, electronic         \\


ZO                  & OATS, electronic           \\

ZR                  & ROUGH RICE, electronic     \\

ZU                  & CRUDE OIL, electronic      \\
ZW                  & WHEAT, electronic          \\
ZZ                  & LEAN HOGS, electronic     \\ \hline

\multicolumn{2}{l}{{\underline{\textbf{Equities (EQ)}}}}\\

EN                  & NASDAQ, MINI               \\

ES                  & S\&P 500, MINI           \\

MD                  & S\&P 400 (Mini electronic)  \\
SC                  & S\&P 500, composite      \\
SP                  & S\&P 500, day session    \\

XX                  & DOW JONES STOXX 50         \\
YM                  & Mini Dow Jones (\$5.00)    \\ \hline
\multicolumn{2}{l}{{\underline{\textbf{Fixed Income (FI)}}}}\\
DT                  & EURO BOND (BUND)        \\
FB                  & T-NOTE, 5yr composite    \\
TY                  & T-NOTE, 10yr composite    \\
UB                  & EURO BOBL                \\
US                  & T-BONDS, composite       \\ \hline
\multicolumn{2}{l}{{\underline{\textbf{Foreign Exchange (FX)}}}}\\
AN & AUSTRALIAN \$\$, composite    \\


DX & US DOLLAR INDEX            \\
FN & EURO, composite            \\
JN & JAPANESE YEN, composite    \\

SN & SWISS FRANC, composite      \\ \bottomrule
\end{tabular}
\end{minipage}
\begin{minipage}{.5\linewidth}
\centering
\caption{Assets in $\mathcal{I}_{ts}$ for zero-shot experiments and additional assets for $\mathcal{I}$ in few-shot experiments. For zero-shot experiments $|\mathcal{I}_{ts}|=20$ and for few-shot experiments  $|\mathcal{I}|=50$.\label{tab:dataset-zeroshot}}
\begin{tabular}{ll}
\midrule
\textbf{Identifier} & \textbf{Description}        \\ \midrule
\multicolumn{2}{l}{{\underline{\textbf{Commodities (CM)}}}}\\
GI                  & GOLDMAN SAKS C. I.         \\
JO                  & ORANGE JUICE               \\
KC                  & COFFEE                     \\
KW                  & WHEAT, KC                  \\
NR                  & ROUGH RICE                 \\
ZG                  & GOLD, electronic           \\
ZH                  & HEATING OIL, electronic    \\
ZK                  & COPPER, electronic         \\
ZL                  & SOYBEAN OIL, electronic    \\
ZN                  & NATURAL GAS, electronic    \\
ZP                  & PLATINUM, electronic       \\
ZT                  & LIVE CATTLE, electronic    \\
 \hline
\multicolumn{2}{l}{{\underline{\textbf{Equities (EQ)}}}}\\
CA                  & CAC40 INDEX                \\
ER                  & RUSSELL 2000, MINI         \\
LX                  & FTSE 100 INDEX             \\
NK & NIKKEI INDEX                \\
XU                  & DOW JONES EUROSTOXX50      \\
\hline
\multicolumn{2}{l}{{\underline{\textbf{Foreign Exchange (FX)}}}}\\
BN & BRITISH POUND, composite   \\
CN & CANADIAN \$\$, composite      \\
MP & MEXICAN PESO               \\
\bottomrule
\end{tabular}
\end{minipage}

\begin{minipage}{.45\linewidth}
\centering

\end{minipage}

\end{table}

\end{appendices}

\end{document}